\documentclass[twocolumn,preprintnumbers,amsmath,amssymb,showpacs]{revtex4}
\usepackage{epsfig}
\usepackage{color}
\usepackage{subfigure}
\usepackage{epstopdf}

\begin{document}
\allowdisplaybreaks
\setlength{\voffset}{1.0cm}
\title{Current response, structure factor and hydrodynamic quantities of a two- and three-dimensional Fermi gas from the operator-product expansion}
\author{Johannes Hofmann}
\email{j.b.hofmann@damtp.cam.ac.uk}
\affiliation{Department of Applied Mathematics and Theoretical Physics, University of Cambridge, Centre for Mathematical Sciences,
Cambridge CB3 0WA, United Kingdom}
\date{\today}
\begin{abstract}
We apply the operator-product expansion to determine the asymptotic form of the current response of a Fermi gas in two and three dimensions. The leading-order term away from the one-particle peak is proportional to a quantity known as the contact, the coefficient of which is determined exactly. We also calculate the dynamic structure factor and the high-frequency tails of the spectral viscosities as a function of the scattering length. Our results are used to derive certain sum rules for the viscosities.
\end{abstract}
\maketitle

\section{INTRODUCTION}
Two-component fermionic quantum gases belong to the most versatile systems that can be used to study the many-body physics of fermions \cite{1,2}. The simplest Lagrangian to describe such a system models the $s$-wave interaction between the constituents by a contact interaction (see, for example, Ref.~\cite{3}):
\begin{equation}
{\cal L} = \psi_\sigma^\dagger \left(i \partial_t + \frac{\nabla^2}{2m}\right) \psi_\sigma - \lambda \psi_\uparrow^\dagger \psi_\downarrow^\dagger \psi_\downarrow^{} \psi_\uparrow^{} , \label{eq:1}
\end{equation}
where a summation over spin indices $\sigma = \uparrow, \downarrow$ is implied. In experiments with cold fermionic atoms, it is possible to confine the atomic cloud in two- and three-dimensional traps and also to tune the interaction strength to a regime where the components are strongly interacting. The theoretical analysis of a Fermi gas in a strongly interacting limit is intricate since many theoretical methods rely on a perturbative expansion or neglect fluctuations and are thus not applicable.

Over the past few years, a lot of work has been stimulated by the observation that some properties of a Fermi gas can be described by \emph{universal relations}. They are called universal because they hold over a wide range of the system's parameters such as the temperature, the density, the precise form \emph{or the strength} of the interaction, and so on. A first set of such relations was presented by Tan \cite{4,5,6}, who among other things derived expressions for the momentum distribution, the energy, and various other thermodynamic quantities, and are hence often called Tan relations. The scale of the Tan relations is set by a quantity known as the contact $C$, which is determined by the many-particle properties of the system. It is defined as $C = \int {\rm d}^d {\bf R} \, \langle {\cal C}({\bf R}) \rangle$, where ${\cal C}({\bf R}) = m^2 \lambda^2 \psi_\uparrow^\dagger \psi_\downarrow^\dagger \psi_\downarrow^{} \psi_\uparrow^{} ({\bf R})$ is the contact density operator and $d = 2,3$ denotes the number of space dimensions. An operational definition of the contact can be given as the magnitude of the large-momentum tail of the gas: $C = \lim_{k \rightarrow \infty} n(k) k^4$. Apart from the dependence on the contact, the universal relations can be inferred from the few-body physics and can be calculated in closed analytic form in many cases. Only the value of the contact has to be taken from numerical calculations or experiments. Some of the Tan relations have been tested experimentally, and good agreement with the theoretical predictions has been found \cite{7}. Since their proposal, many other exact relations for fermionic as well as bosonic systems have been derived (cf., for example, Refs.~\cite{15, 16, 11, 12, 14, 8, 9, 10, 13, 16a}). An overview of various universal relations and recent experimental developments can be found in Ref.~\cite{17}.

A field-theoretical derivation of universal relations was put forward by Braaten, Kang, and Platter \cite{18, 19}, who used the operator-product expansion (OPE) to rederive the Tan relations \cite{4, 5, 6}. The OPE expresses the product of two local operators ${\cal O}_i^{(a)}$ and ${\cal O}_j^{(b)}$ at different points in space and time in terms of local operators ${\cal O}^{(k)}_{\boldsymbol{\alpha}}$, where ${\boldsymbol{\alpha}} = (\alpha_1, \alpha_2, \ldots)$ denotes a set of indices \cite{20}:
\begin{equation}
 {\cal O}^{(a)}_i({\bf R} + {\bf r}, T + t) {\cal O}_j^{(b)}({\bf R}, T) = \sum_{k, \boldsymbol{\alpha}} c^{(k)}_{ij\boldsymbol{\alpha}}({\bf r}, t) {\cal O}^{(k)}_{\boldsymbol{\alpha}}({\bf R}, T) . \label{eq:2}
\end{equation}
Note that, on the right-hand side, the dependence on the separation $({\bf r}, t)$ is carried by the \emph{Wilson coefficients} $c^{(k)}_{ij\boldsymbol{\alpha}}$, which can be calculated exactly for simple operators. If an observable can be expressed as the expectation value of an operator product involving operators at different space-time points, we can use the OPE to determine its leading-order behavior for small ${\bf r}$ and $t$ (or for large values of momentum ${\bf q}$ and energy $\omega$ for the Fourier transform). The advantage of the OPE is that it provides a powerful framework to derive additional universal relations. This was exploited by the authors of \cite{19} to derive universal results for radio-frequency spectroscopy measurements \cite{11}. Son and Thompson studied the asymptotic behavior of the dynamic structure factor $S({\bf q}, \omega)$ at unitarity at large ${\bf q}$ and $\omega$ as well as the large-frequency tail of $S$ \cite{12}. The result for the large-frequency tail in \cite{12} was augmented by Goldberger and Rothstein, who calculated the leading-order correction due to finite scattering length \cite{21}. Some of the above results have also been obtained by other authors using different methods \cite{14, 15, 16}. While most of the above work was carried out for a three-dimensional system at or close to unitarity, it is possible to derive universal relations for any strength of the interaction or any other dimension in the same way.

In this work, we apply the OPE to derive the current response function of a Fermi gas in two and in three dimensions. It is defined as
\begin{equation}
G_{ij}^{ \sigma \sigma'}({\bf x} - {\bf y}, t) = i \Theta(t) \langle [j_{i}^{\sigma}({\bf x}, t), j_{j}^{\sigma'}({\bf y}, 0)] \rangle , \label{eq:3}
\end{equation}
where $j_i^\sigma({\bf x}, t) = - (i/2 m) (\psi_\sigma^\dagger\, \overset{\leftrightarrow}{\partial_i} \psi_\sigma) \equiv - (i/2 m) (\psi_\sigma^\dagger \partial_i \psi_\sigma \linebreak  - \partial_i \psi_\sigma^\dagger \psi_\sigma)$ is the current operator. The Fourier transform of Eq.~(\ref{eq:3}) can be split into a transverse and a longitudinal part:
\begin{equation}
G_{ij}^{\sigma \sigma'}({\bf q}, \omega) = G_L^{\sigma \sigma'}({\bf q}, \omega) \frac{q_i q_j}{q^2} + G_T^{\sigma \sigma'}({\bf q}, \omega) \left(\delta_{ij} - \frac{q_i q_j}{q^2}\right) .
\end{equation}
The longitudinal part is related to the superfluid fraction and the transverse part to the normal fraction of the gas. Thus, the current response function is a central quantity in the study of superfluids. The longitudinal component is related to the structure factor and can be determined with methods that measure the density response, such as two-photon Bragg spectroscopy \cite{22}.

Recently, there have been several proposals to probe the transverse component of the current response using artificial gauge fields \cite{23, 24, 25}. Artificial gauge fields are induced by either rotating the atomic cloud or placing it in a specific laser setup. In the latter case, incident laser beams couple the ground state of the atoms to an excited state. The effective Lagrangian of such a system describes charged particles coupled to a U(1) gauge field. The amplitude of this gauge field typically depends on the spatial or temporal variation of the Rabi frequency or the detuning of the laser from the resonance frequency. For example, a local detuning of the laser field could induce a rapidly oscillating gauge field. For further references, see~\cite{26, 27}. One way to measure the imaginary part of the current response is to determine how much energy $W$ is absorbed by the system if the gauge field is switched on for a short time:
\begin{align}
W &= 2 \int_0^\infty \frac{{\rm d}\omega}{2 \pi} \int \frac{{\rm d}^d{\bf q}}{(2 \pi)^d} \, A_{\sigma}^i(- {\bf q}, - \omega) \, \omega \, {\rm Im} \, G_{ij}^{\sigma \sigma'}({\bf q}, \omega) \nonumber \\
&\quad \times A_{\sigma'}^j({\bf q}, \omega). 
\end{align}
For a two-dimensional gas, there also exists a proposal to determine the current response optically, by measuring the phase shift of a laser beam that crosses the atomic cloud \cite{25}. While artificial gauge fields have only been realized for Bose gases so far there is no reason why this should not also be possible for fermionic systems in the near future. It is in anticipation of such experiments that we calculate the current response. In particular, since it should be possible to engineer gauge fields that couple differently to each spin species, we consider both the spin-symmetric and antisymmetric current response $G_{ij}^{s/a} = G_{ij}^{\sigma\sigma} \pm G_{ij}^{\sigma -\sigma}$, where a summation over spin indices is implied [corresponding to gauge fields that couple symmetrically and antisymmetrically to the current: ${\bf A} \cdot ({\bf j}_\uparrow \pm {\bf j}_{\downarrow})$].

This paper is structured as follows: In Sec.~\ref{sec:2}, we describe how to calculate the Wilson coefficients of the current correlator. This is done using a matching procedure where the matrix element of Eq.~(\ref{eq:2}) is evaluated with respect to selected few-body states. We choose a one-particle state to determine the Wilson coefficients of the bilinear operators (Sec.~\ref{sec:2a}) and then a two-particle state for the contact operator (Sec.~\ref{sec:2b}). Section~\ref{sec:3} presents the results of this calculation. Expressions for both the spin-symmetric and the antisymmetric current response are presented and discussed in Sec.~\ref{sec:3a}. The U(1) symmetry of the system implies a Ward identity that relates the longitudinal part of the current correlator to the dynamic structure factor. We use this to calculate the asymptotic form of the dynamic structure factor in two and in three dimensions for any value of the scattering length in Sec.~\ref{sec:3b}. In \cite{14}, Taylor and Randeria derived expressions for the large-frequency tail of the spectral shear and bulk viscosity $\eta(\omega)$ and $\zeta(\omega)$. This paper reproduces their results using a field-theoretical framework and generalizes them to finite scattering length. We are also able to provide the viscosity tails for a two-dimensional Fermi gas. We conclude with a summary in Sec.~\ref{sec:4}. The paper contains an appendix that presents details of the calculations in Sec.~\ref{sec:2}.

\section{OPERATOR PRODUCT EXPANSION}\label{sec:2}

While we would like to obtain the OPE for the retarded response function $G_{ij}^{\sigma \sigma'}({\bf q}, \omega)$, it is more convenient to calculate the time-ordered response, which coincides with the retarded response up to a correction of order ${\cal O}(2 e^{- \beta \omega})$ \cite{28}:
\begin{align}
&G_{ij}^{\sigma \sigma'}({\bf q}, \omega) = i \int {\rm d}t \, e^{i (\omega + i \varepsilon) t} \int {\rm d}^d{\bf r} \int {\rm d}^d{\bf R} \, e^{- i {\bf q} \cdot {\bf r}} \nonumber \\
&\quad \times \left\langle {\cal T} j_{i}^{\sigma}\left({\bf R} + \frac{{\bf r}}{2}, t\right) \, j_{j}^{\sigma'}\left({\bf R} - \frac{{\bf r}}{2}, 0\right) \right\rangle + {\cal O}(2 e^{- \beta \omega}). \label{eq:6}
\end{align}
${\cal T}$ denotes the time-ordered product. The $i \epsilon$ term in the Fourier transform has been included to shift the poles in the resulting expressions of the response functions \cite{28}. Using the time-ordered response allows us to apply standard diagrammatic methods to calculate the Wilson coefficients. In this section, we derive the OPE of the operator
\begin{align}
&A_{ij}^{\sigma \sigma'}({\bf q}, \omega) = \int {\rm d}t \int d^d{\bf r} \int {\rm d}^d{\bf R} \, e^{i \omega t - i {\bf q} \cdot {\bf r}} \nonumber \\*
&\quad \times {\cal T} j_{i}^{\sigma}\left({\bf R} + \frac{{\bf r}}{2}, t\right) \, j_{j}^{\sigma'}\left({\bf R} - \frac{{\bf r}}{2}, 0\right) . \label{eq:7}
\end{align}

The OPE is an operator identity, that is, it holds if we take the expectation value of Eq.~(\ref{eq:7}) with respect to any state. Hence, in order to determine the Wilson coefficient $c^{(k)}$ of an operator ${\cal O}^{(k)}$, we can choose a simple few-body state for which $\langle {\cal O}^{(k)} \rangle \neq 0$ and match the expectation values on both sides of Eq.~(\ref{eq:2}). Inserting Eq.~(\ref{eq:2}) in Eq.~(\ref{eq:7}) yields
\begin{align}
&A^{\sigma \sigma'}_{ij}({\bf q}, \omega) = \sum_{k, \boldsymbol{\alpha}} \frac{1}{\omega^{\Delta_k/2 - d/2}} \, c_{ij\boldsymbol{\alpha}}^{(k)} \left(\frac{q^2}{2 m \omega}, \frac{a^{-1}}{\sqrt{m \omega}}\right) \nonumber \\
&\quad \times \int {\rm d}^d{\bf R} \, {\cal O}^{(k)}_{\boldsymbol{\alpha}}({\bf R}), \label{eq:8}
\end{align} 
where $\Delta_k$ denotes the scaling dimension of the operator ${\cal O}^{(k)}_{\boldsymbol{\alpha}}$ as defined by ${\cal O}_{\boldsymbol{\alpha}}^{(k)}(\lambda {\bf x}, \lambda^2 t) = \lambda^{- \Delta_k} {\cal O}_{\boldsymbol{\alpha}}^{(k)}({\bf x}, t)$. We suppress the dependence of the Wilson coefficients on the spin indices on the right-hand side of Eq.~(\ref{eq:8}). The operators ${\cal O}_{\boldsymbol{\alpha}}^{(k)}$ can be composed of the field operators $\psi_\sigma$ and their adjoints $\psi_\sigma^\dagger$ as well as derivative operators acting on those fields. We will only consider operators that conserve the particle number, i.e., contain the same number of fields and their adjoints, and that are symmetric in the spin indices. Equation~(\ref{eq:8}) shows that the high-frequency behavior of the current response is governed by the operators with lowest scaling dimension.
Those operators are 
\begin{alignat}{2}
n &= \psi_\sigma^\dagger \psi_\sigma & &(\Delta_n = d), \label{eq:9} \\
{\bf j} &= - \frac{i}{2 m} \psi_\sigma^\dagger \overset{\leftrightarrow}{\nabla}\psi_\sigma &  &(\Delta_{\bf j} = d + 1), \\
{\cal O}_{i j}^{(3)} &= \frac{1}{2 m} \psi_\sigma^\dagger \overset{\leftrightarrow}{\partial_i} \overset{\leftrightarrow}{\partial_j} \psi_\sigma & &(\Delta_{{\cal O}^{(3)}} = d + 2), \\
{\cal O}^{(4)} &= \psi_\sigma^\dagger \Bigl(i \partial_t + \frac{\nabla^2}{2 m}\Bigr) \psi_\sigma \quad & &(\Delta_{{\cal O}^{(4)}} = d + 2) , \label{eq:12}
\end{alignat}
and the contact density operator
\begin{equation}
{\cal C} = m^2 \lambda^2 \psi_\uparrow^\dagger \psi_\downarrow^\dagger \psi_\downarrow^{} \psi_\uparrow^{} \quad {\rm with} \quad  \Delta_{\cal C} = 4.
\end{equation}
We use the shorthand ${\cal \psi}^\dagger \overset{\leftrightarrow}{\partial} \psi = {\cal \psi}^\dagger \partial \psi - \partial {\cal \psi}^\dagger \psi$.

We proceed as follows: First, we summarize the Feynman rules of the theory in Eq.~(\ref{eq:1}) and match the bilinear operators~(\ref{eq:9})--(\ref{eq:12}) using a one-particle state (for which the matrix element of the contact density vanishes). Then, we determine the Wilson coefficient of the contact using a two-particle state and the results of Sec.~\ref{sec:2a}.

\subsection{Feynman rules}\label{sec:2o}

\begin{figure}[t!]
\begin{center}
\subfigure[]{\scalebox{0.6}{\epsfig{file=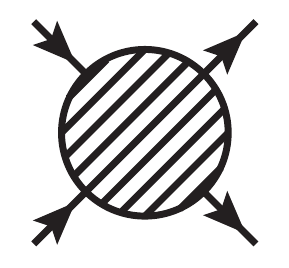}}\label{fig:1a}}\qquad
\subfigure[]{\scalebox{0.6}{\epsfig{file=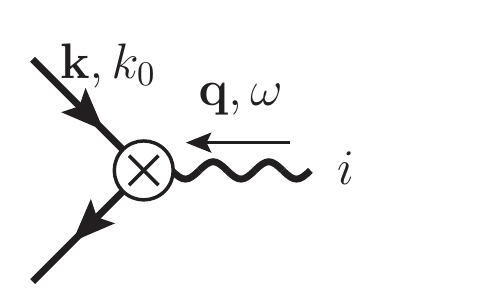}}\label{fig:1b}}
\caption{(a) Feynman diagram for the two-particle scattering amplitude. (b) Current insertion on a fermion line.} 
\label{fig:1}
\end{center}
\end{figure}
The matrix element of an operator can be expressed pictographically in terms of Feynman diagrams. This section lists the Feynman rules that are necessary to evaluate those diagrams.

Every line is assigned an energy and a momentum where we impose momentum and energy conservation at each vertex. The external energy and momentum are constrained by the state. After applying the Feynman rules, we integrate over each undetermined loop momentum ${\bf k}$ and energy $k_0$ using the measure $\int {\rm d}^d{\bf k} \, dk_0 /(2 \pi)^{d + 1}$. Each internal line of a particle with energy $k_0$ and momentum ${\bf k}$ contributes a factor $i/(k_0 - E_{\bf k} + i \varepsilon)$ to the integrand. The dashed blob [Fig.~\ref{fig:1a}] denotes the two-particle scattering amplitude and gives a term $i {\cal A}(k_0, {\bf k})$, where $k_0$ and ${\bf k}$ are the sum of the energies and the momenta of the incoming fermions, respectively. We have
\begin{equation}
{\cal A}({\bf k}, k_0) = \frac{4 \pi}{m} \frac{1}{- a_{\rm 3D}^{-1} + \sqrt{- m (k_0 - {\bf k}^2/4m) - i \varepsilon}}
\end{equation}
in three dimensions (3D) and
\begin{equation}
{\cal A}({\bf k}, k_0) = \frac{2 \pi}{m} \frac{1}{\ln a_{\rm 2D} \sqrt{- m (k_0 - {\bf k}^2/4m) - i \varepsilon}} \label{eq:15}
\end{equation}
in two dimensions (2D). In the following, we abbreviate ${\cal A}(0, {\bf 0}) \equiv {\cal A}$. The strength of the interaction is given by the 2D and 3D scattering lengths $a_{\rm 2D}$ and $a_{\rm 3D}$, respectively. ${\cal A}(k_0, {\bf k})$ can be obtained by summing all ladder diagrams that contribute to the two-atom Green's function. The bare coupling constant $\lambda$ is related to ${\cal A}$ by the renormalization condition
\begin{equation}
{\cal A}^{-1} = - \frac{1}{\lambda} - \int \frac{{\rm d}^d{\bf k}}{(2 \pi)^d} \frac{1}{2 E_{\bf k}}. \label{eq:16}
\end{equation}
This condition ensures that the two-particle scattering amplitude reproduces the leading order in the effective range expansion of the quantum-mechanical $s$-wave scattering phase shift, $k \cot \delta(k) = - a_{\rm 3D}^{-1}$ in 3D \cite{3, 29} and $\cot \delta(k) = (1/\pi) \ln k^2 a_{\rm 2D}^2$ in 2D \cite{30, 31}, respectively.
 
We denote the insertion of a current operator by a crossed circle attached to a wiggly line. A current insertion with energy $\omega$ and momentum ${\bf q}$ that is attached to a fermion line with energy $k_0$ and momentum ${\bf k}$ is depicted in Fig.~\ref{fig:1b}. It contributes a factor
\begin{equation}
\frac{1}{2 m} (2 k_i + q_i)
\end{equation}
to the integrand. The Feynman rules of the operators~(\ref{eq:9})--(\ref{eq:12}) are listed in Eqs.~(\ref{eq:19})--(\ref{eq:22}).

\subsection{One-particle state}\label{sec:2a}

\begin{figure}[t!]
\begin{center}
\scalebox{0.6}{\epsfig{file=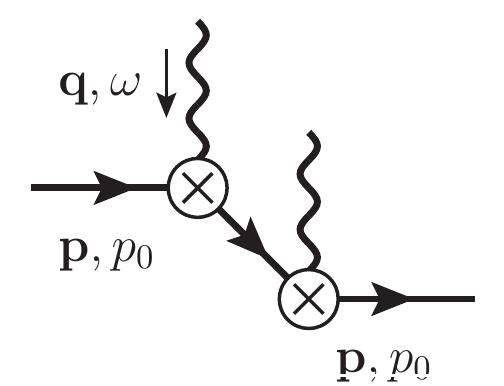}}
\scalebox{0.6}{\epsfig{file=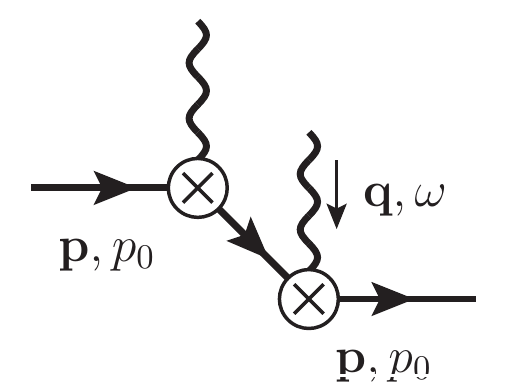}}
\caption{Diagrams that contribute to the expectation value of $A_{ij}^{\sigma \sigma'}({\bf q}, \omega)$ [Eq.~(\ref{eq:7})] between a one-particle state with momentum ${\bf p}$ and energy $p_0$.} 
\label{fig:2}
\end{center}
\end{figure}
In order to determine the Wilson coefficients of the bilinear operators, we evaluate the matrix element of $A_{ij}^{\sigma \sigma'}$ [Eq.~(\ref{eq:7})] with respect to a one-particle state with momentum ${\bf p}$ and energy $p_0$. Note that we do not enforce the on-shell condition $p_0 = {\bf p}^2/2m$. The relevant Feynman diagrams are depicted in Fig.~\ref{fig:2}. Their value is:
\begin{align}
&\langle {\bf p}, p_0 | A_{ij}^{\uparrow \uparrow}({\bf q}, \omega) | {\bf p}, p_0 \rangle = \frac{i/4m^2}{p_0 + \omega - E_{\bf p + q}} \nonumber \\
&\quad \times (2 p_i + q_i) (2 p_j + q_j) + (\omega, {\bf q} \rightarrow - \omega, - {\bf q}), \label{eq:18}
\end{align}
where $E_{\bf p} = {\bf p}^2/2m$. The same holds for $A_{ij}^{\downarrow \downarrow}$. The expectation values of the bilinear operators are
\begin{align}
\langle {\bf p}, p_0 | n | {\bf p}, p_0 \rangle &= 1 , \label{eq:19} \\
\langle {\bf p}, p_0 | j_\alpha | {\bf p}, p_0 \rangle &= p_\alpha/m , \\
\langle {\bf p}, p_0 | {\cal O}_{\alpha \beta}^{(3)} | {\bf p}, p_0 \rangle &= - 2 p_\alpha p_\beta/m \quad {\rm and} \\
\langle {\bf p}, p_0 | {\cal O}^{(4)} | {\bf p}, p_0 \rangle &= p_0 - E_{\bf p}. \label{eq:22}
\end{align}
They match the first terms in the expansion of Eq.~(\ref{eq:18}) in terms of $p_0$ and ${\bf p}$. This yields the Wilson coefficients:
\begin{align}
 c_{ij}^n &= \frac{i/4m}{m \omega - {\bf q}^2/2} q_i q_j + (\omega \rightarrow - \omega) ,  \label{eq:23} \\
 c_{ij\alpha}^{\, {\bf j}} &= \frac{i/4}{(m \omega - {\bf q}^2/2)^2} q_\alpha q_i q_j + \frac{i/2}{m \omega - {\bf q}^2/2} \nonumber \\
&\quad \times \left(\delta_{i\alpha} q_j + \delta_{j\alpha} q_i\right)  - (\omega \rightarrow - \omega) , \\
 c_{ij\alpha\beta}^{(3)} \delta_{\alpha\beta} &= \left[\frac{- i q^2/8}{(m \omega - q^2/2)^3} - \frac{i/2}{(m \omega - q^2/2)^2} \right] q_i q_j \nonumber \\
&\quad - \frac{i/2}{m \omega - q^2/2} \delta_{ij} + (\omega \rightarrow - \omega) \label{eq:25}
\end{align}
and
\begin{equation}
 c_{ij}^{(4)} = \frac{-i/4}{(m \omega - q^2/2)^2} q_i q_j + (\omega \rightarrow - \omega). \label{eq:26}
\end{equation}
For an isotropic system, we have $\langle {\cal O}_{\alpha\beta}^{(3)} \rangle~=~(2/md) \linebreak \langle \psi_\sigma^\dagger \nabla^2 \psi_\sigma\rangle \delta_{\alpha\beta}$, hence, only the contraction $c_{ij\alpha\beta}^{(3)} \delta_{\alpha\beta}$ will be relevant in the following. Furthermore, we have $\langle {\bf j} \rangle = 0$ and can apply the Heisenberg equation of motion of the operator ${\cal O}^{(4)}$: $\langle {\cal O}^{(4)} \rangle = (2/m^2\lambda) \langle {\cal C} \rangle$. The OPE allows us to express the expectation value of Eq.~(\ref{eq:7}) as follows:
\begin{align}
&\langle A_{ij}^{\sigma \sigma'}({\bf q}, \omega) \rangle = c_{ij}^n({\bf q}, \omega) \, N + c_{ij}^{\cal H}({\bf q}, \omega) \, H \nonumber  \\*
&\quad + c_{ij}^{\cal C}({\bf q}, \omega) \, C + {\cal O}\left(\frac{1}{\omega^{3-d/2}}\right) . \label{eq:27}
\end{align}
Here, $N = \int {\rm d}^d{\bf R} \, \langle n({\bf R}) \rangle$ and $H = \int {\rm d}^d{\bf R} \, \langle {\cal H}({\bf R}) \rangle$, where ${\cal H}$ denotes the Hamiltonian density
\begin{equation}
 {\cal H}({\bf R}) = \frac{1}{2 m} \nabla \psi_\sigma^\dagger \nabla \psi_\sigma ({\bf R}) + \lambda \psi_\uparrow^\dagger \psi_\downarrow^\dagger \psi_\downarrow^{} \psi_\uparrow^{} ({\bf R}),
\end{equation}
which has the Wilson coefficient
\begin{equation}
 c_{ij}^{\cal H}({\bf q}, \omega) = - \frac{4}{d} \delta^{}_{\alpha\beta} c_{ij\alpha\beta}^{(3)}({\bf q}, \omega). \label{eq:29}
\end{equation}
From Eq.~(\ref{eq:27}), we see that the terms proportional to $N$ and $H$ only contribute a term $\sim \delta(m \omega - q^2/2)$ to the imaginary part of $G_{ij}^{\sigma\sigma'}$. Away from this one-particle peak, the leading order is proportional to the contact. In the next section, we proceed to calculate the Wilson coefficient $c_{ij}^{\cal C}({\bf q}, \omega)$ of the contact operator.

\subsection{Two-particle state}\label{sec:2b}

To determine the Wilson coefficient of the contact operator, we choose a two-particle state where both particles have zero energy and momentum. This way, the matrix element of other two-body operators with higher scaling dimension than the contact, which involve additional derivatives of the field, are zero. We only have to subtract the matrix elements of the one-particle operators with respect to the two-particle state. We do, however, introduce additional infrared divergences when calculating single diagrams, which we need to subtract before we can evaluate the finite part. In particular, in two dimensions, this makes the calculations somewhat lengthy. Of course, the overall result is finite and the divergent parts cancel when summing all diagrams.
\begin{figure}[t!]
\begin{center}
\scalebox{0.6}{\epsfig{file=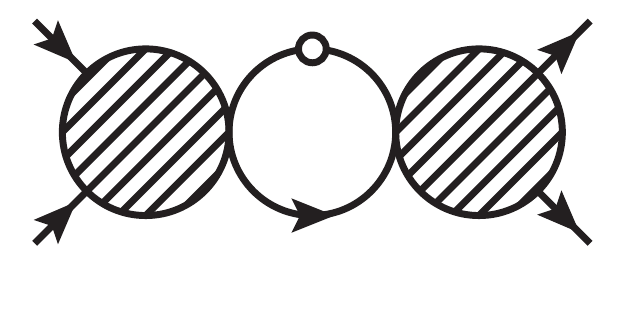}}
\caption{Insertion of a bilinear operator on an internal line. The insertion on an external line matches the insertion of two current operators on the same line and need not be considered.} 
\label{fig:3}
\end{center}
\end{figure}

We start by calculating the insertion of the bilinear operators on an internal line as shown in Fig.~\ref{fig:3}. To obtain a finite expression, we introduce an IR and a UV cutoff $\eta$ and $\Lambda$, respectively:
\begin{align}
&\left. \left\langle 0, \uparrow; 0, \downarrow \left| \, n \, \right| 0, \uparrow; 0, \downarrow \right\rangle \right|_{{\rm Fig.} \, \ref{fig:3}} \equiv \langle n \rangle\nonumber \\*
&\quad=  m^2 {\cal A}^2 \times \left\{ \begin{matrix}
                 i/4 \pi \eta & d = 3 \\
                 - 1/2 \pi \eta^2 & d = 2
                 \end{matrix} \right. , 
\end{align}
\begin{align}
&\left. \langle 0, \uparrow; 0, \downarrow | \, {\cal H} \, | 0, \uparrow; 0, \downarrow \rangle \right|_{{\rm Fig.} \, \ref{fig:3}} \equiv \langle {\cal H} \rangle \nonumber \\
&\quad= \frac{m^2 {\cal A}^2}{m} \times \left\{ \begin{matrix}
           a_{\rm 3D}^{-1}/4\pi & d = 3 \\
           \frac{1}{2 \pi} \left(- \frac{1}{2} \ln a_{\rm 2D}^2 \eta^2 - \frac{1}{2} + \frac{i \pi}{2}\right) & d = 2
           \end{matrix} \right. 
\end{align}
and
\begin{align}
&\left. \langle 0, \uparrow; 0, \downarrow | {\cal O}^{(4)} | 0, \uparrow; 0, \downarrow \rangle \right|_{{\rm Fig.} \, \ref{fig:3}}  \equiv \langle {\cal O}^{(4)} \rangle\nonumber \\
&\quad= - \frac{2 m^2 {\cal A}^2}{m} \left\{ \begin{matrix}
                 \Lambda/2 \pi^2 & d = 3 \\
                 \frac{1}{2 \pi} \left(\frac{1}{2} \ln \frac{\Lambda^2}{\eta^2} + \frac{i \pi}{2}\right) & d = 2
                 \end{matrix} \right. .      	
\end{align}
The matrix element of the contact operator is
\begin{equation}
\left\langle 0, \uparrow; 0, \downarrow \left| \, {\cal C} \, \right| 0, \uparrow; 0, \downarrow \right\rangle \equiv \langle {\cal C} \rangle =  m^2 {\cal A}^2
\end{equation}
both in two and in three space dimensions. To keep our notation concise, we abbreviate the matrix elements of the various operators as defined in the previous equations.

\begin{figure}[t!]
\begin{center}
\subfigure[tight][]{\scalebox{0.6}{\epsfig{file=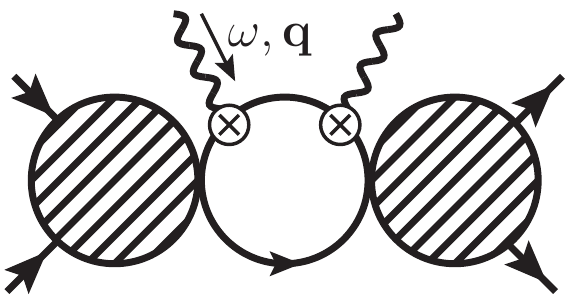}}\label{fig:4a}}\vspace{0.1cm}
\subfigure[tight][]{\scalebox{0.6}{\epsfig{file=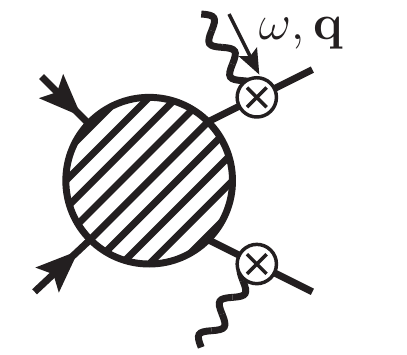}}
\scalebox{0.6}{\epsfig{file=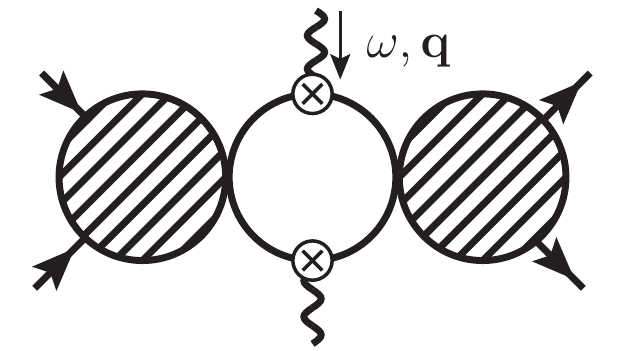}}\label{fig:4b}}\vspace{0.1cm}
\scalebox{0.6}{\epsfig{file=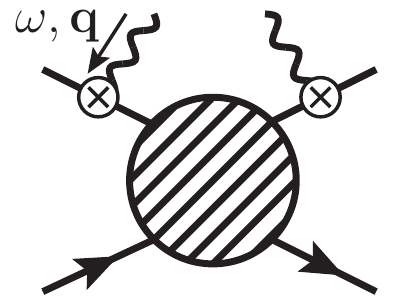}}
\scalebox{0.6}{\epsfig{file=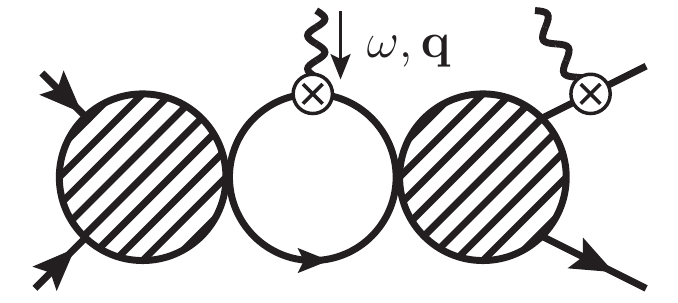}}
\subfigure[tight][]{\scalebox{0.6}{\epsfig{file=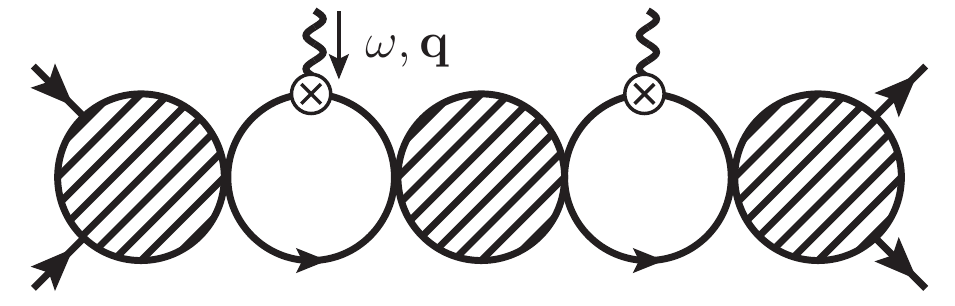}}\label{fig:4c}}
\caption{Diagrams that contribute to the matrix elements of $A_{ij}^{\sigma \sigma'}({\bf q}, \omega)$ [Eq.~(\ref{eq:7})] between the two-particle state.}
\label{fig:4}
\end{center}
\end{figure}

The relevant diagrams that contribute to the matrix elements of Eq.~(\ref{eq:7}) are shown in Fig.~\ref{fig:4}. The insertion of two current operators on an external line has already been considered in the previous section and is matched by bilinear insertion on external lines. Figures~\ref{fig:4a} and~\ref{fig:4b}, respectively, contribute to only the diagonal and off-diagonal components of $\langle A_{ij}^{\sigma \sigma'} \rangle$ in spin space, whereas Fig.~\ref{fig:4c} contributes to all components. The matrix elements of the bilinear operators are diagonal in spin space. Thus, they cancel the divergent parts in Fig.~\ref{fig:4a}. Figures~\ref{fig:4b} and~\ref{fig:4c} are separately finite. In the following, we denote the contribution of Figs.~\ref{fig:4a},~\ref{fig:4b} and~\ref{fig:4c} to the Wilson coefficient of the contact by $C_{ij}^{(1)}$, $C_{ij}^{(2)}$ and $C_{ij}^{(3)}$, respectively, where we absorb the contribution of the bilinear matrix elements into the definition of $C_{ij}^{(1)}$. In short, we have
\begin{equation}
c_{ij}^{{\cal C}} = C_{ij}^{(1)} \frac{\mathbb{I}_{2\times2}}{2} + C_{ij}^{(2)} \frac{\sigma_1}{2} + C_{ij}^{(3)} \frac{\mathbb{I}_{2\times2} + \sigma_1}{4}.
\end{equation}

Using the results of this section, we can give an explicit expression for the coefficient $C_{ij}^{(1)}$:
\begin{align}
&C_{ij}^{(1)} = \frac{i}{2 m^4} \int \frac{{\rm d}^d {\bf k}}{(2 \pi)^d} \frac{1}{4 E_{\bf k}^2} \frac{1}{\omega - E_{\bf k} - E_{\bf k + q} + i \varepsilon} \nonumber \\
&\quad \times (2 k_i + q_i) (2 k_j + q_j) - c_{ij}^n \frac{\langle n \rangle}{\langle {\cal C} \rangle} - c_{ij}^{\cal H}({\bf q}, \omega) \frac{\langle {\cal H} \rangle}{\langle {\cal C} \rangle} \nonumber \\
&\quad  + c_{ij}^{(4)}({\bf q}, \omega) \left[\frac{2}{m^2 \lambda} - \frac{\langle {\cal O}^{(4)} \rangle}{\langle {\cal C} \rangle}\right] + (\omega \rightarrow - \omega). \label{eq:35}
\end{align}
The bilinear contributions cancel the IR divergences of Fig.~\ref{fig:4a} and render the expression~(\ref{eq:35}) finite. In 3D and 2D, Fig.~\ref{fig:4a} has a power-law infrared divergence that is matched by the matrix element of the number operator $n$. In 2D, there are additional logarithmic infrared divergences in both the transverse as well as the longitudinal components. They are canceled by the expectation values of the operators ${\cal H}$ and ${\cal O}^{(4)}$.

The sum of the diagrams in Fig.~\ref{fig:4b} (divided by the matrix element of ${\cal C}$) is equal to
\begin{align}
C_{ij}^{(2)} &= - \frac{i}{m^2 \omega} \int \frac{{\rm d}^d {\bf k}}{(2 \pi)^d} \frac{1}{E_{\bf k}} \frac{1}{\omega - E_{\bf k} - E_{\bf k + q} + i \varepsilon} \nonumber \\
&\quad \times (2 k_i + q_i) (2 k_j + q_j) + \frac{2 i {\cal A}^{-1}}{m^2 \omega} \frac{q_i q_j}{\omega - E_{\bf q}} \nonumber \\
&\quad + (\omega \rightarrow - \omega). \label{eq:36}
\end{align}
In two dimensions, the first term contains a logarithmic infrared-divergent part. It is canceled by the infrared divergence in the two-particle scattering amplitude ${\cal A}$ [cf. Eq.~(\ref{eq:15})].

The last contribution to $c_{ij}^{{\cal C}}$ comes from the three diagrams in Fig.~\ref{fig:4c}. They only contribute to the longitudinal part of the response (cf. the Appendix):
\begin{align}
C_{ij}^{(3)} &= - \frac{i {\cal A}(\omega, {\bf q})}{m^4} \biggl[ \frac{q_i}{\omega - E_{\bf q}} {\cal A}^{-1} + \int \frac{{\rm d}^d {\bf k}}{(2 \pi)^d} \nonumber \\
&\quad \times \frac{1}{2 E_{\bf k}} \frac{1}{\omega - E_{\bf k + q} - E_{\bf k} + i \varepsilon} (2 k_i + q_i)\biggr] \times [i \rightarrow j] \nonumber \\
&\quad + (\omega \rightarrow - \omega). \label{eq:37} 
\end{align}
As before, a logarithmic infrared divergence in 2D is canceled between the two terms and the expression is manifestly finite.

It is possible but cumbersome to evaluate the expressions (\ref{eq:35}), (\ref{eq:36}), and (\ref{eq:37}) in closed analytical form in terms of elementary functions. For details of this calculation we refer to the Appendix. The results of the computation are presented in the next section.

\section{RESULTS}\label{sec:3}

The first part of this section lists and discusses the current response functions of a Fermi gas in two and in three dimensions for large values of the arguments ${\bf q}$ and $\omega$, where $x = q^2/2m\omega$ and $a^{-1}/\sqrt{m \omega}$ are kept fixed. We consider both the response to a gauge field that couples symmetrically as well as antisymmetrically to the different fermion species. The latter corresponds to probing the response of the spin-antisymmetric current ${\bf j}^a =  {\bf j}^\uparrow - {\bf j}^\downarrow$. We denote the spin-symmetric and antisymmetric response function by a superscript $s$ and $a$, respectively. In the remainder of this section, we use the results of Sec.~\ref{sec:3a} to calculate the dynamic structure factor in 3D and 2D. We also determine the asymptotic form of the spectral viscosities introduced in \cite{14} for large values of the frequency, and comment on the derivation of sum rules using the OPE results.

\begin{figure*}[t]
\centering
\subfigure[]{\includegraphics[scale=0.9]{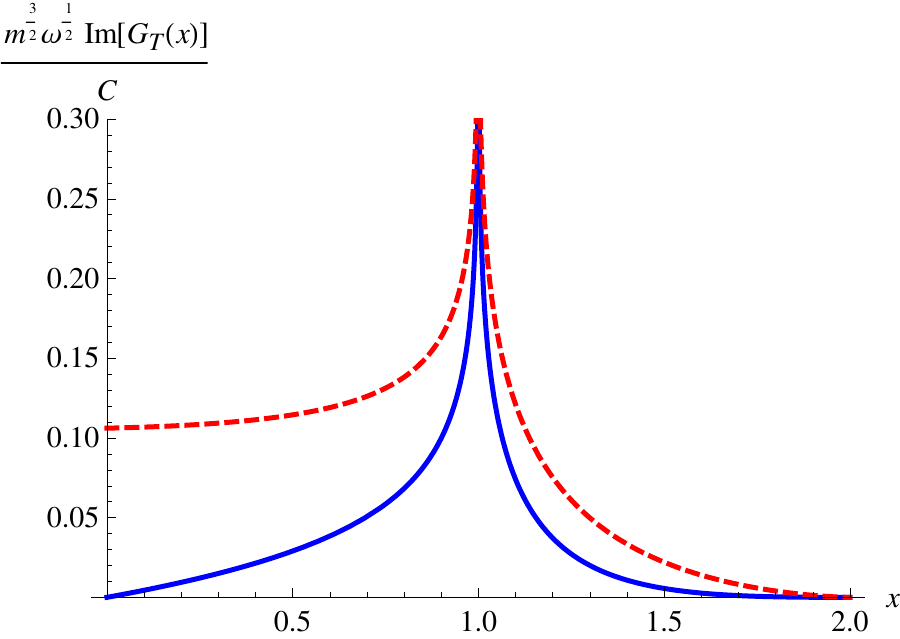}\label{fig:5a}} \qquad
\subfigure[]{\includegraphics[scale=0.9]{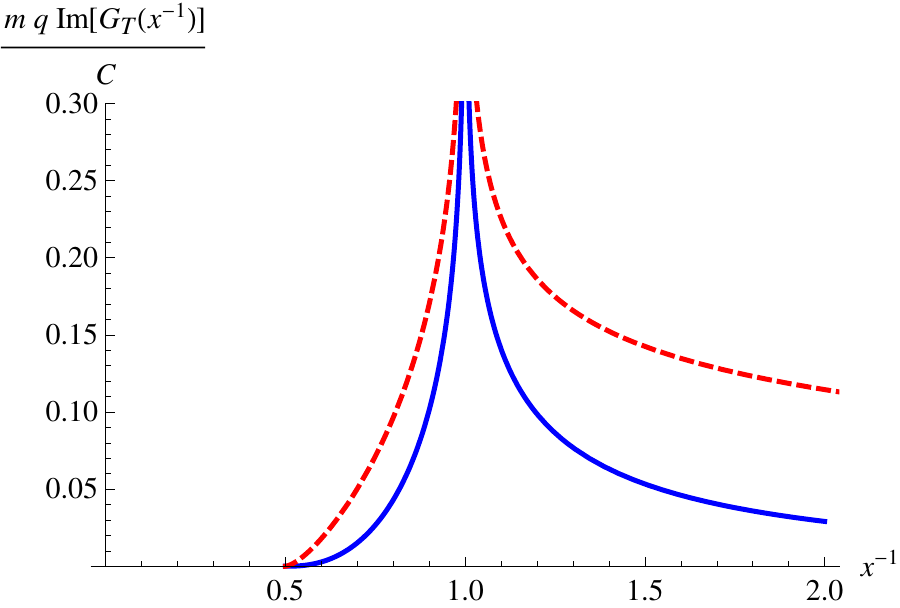}\label{fig:5b}}
\caption{3D: Transverse response function at (a) fixed energy and (b) fixed momentum as a function of the variable $x = q^2/2m\omega$ and $x^{-1}$, respectively. The continuous blue line denotes the spin-symmetric and the red dashed line the spin-antisymmetric response.}
\label{fig:5}
\end{figure*}
\begin{figure*}[t]
\centering
\subfigure[]{\includegraphics[scale=0.9]{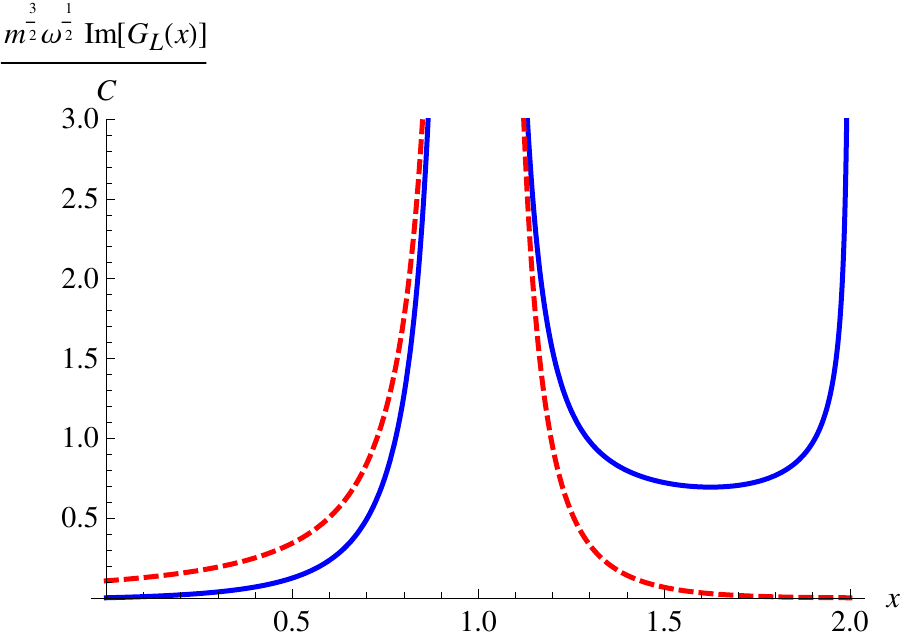}\label{fig:6a}} \qquad
\subfigure[]{\includegraphics[scale=0.9]{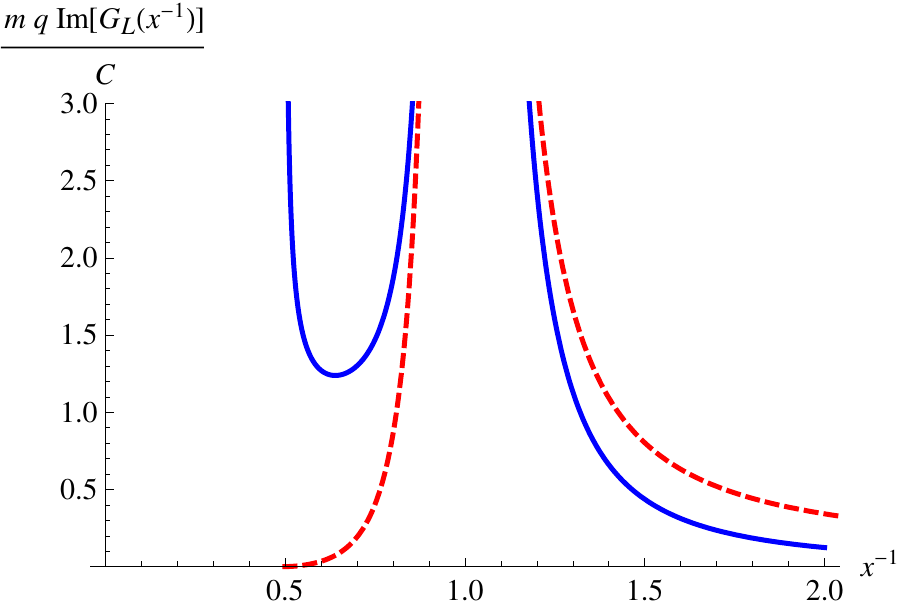}\label{fig:6b}}
\caption{3D: Longitudinal response function at (a) fixed energy and (b) fixed momentum as a function of the variable $x = q^2/2m\omega$ and $x^{-1}$, respectively. The continuous blue line denotes the spin-symmetric and the red dashed line the spin-antisymmetric response. The spin-symmetric response is evaluated at unitarity.}
\label{fig:6}
\end{figure*}

\subsection{Response functions}\label{sec:3a}

As discussed in Sec.~\ref{sec:2b}, $C_{ij}^{(1)}$ [Eq.~(\ref{eq:35})] contributes to the diagonal and $C_{ij}^{(2)}$ [Eq.~(\ref{eq:36})] to the off-diagonal components of the contact's Wilson coefficient in spin space, whereas $C_{ij}^{(3)}$ [Eq.~(\ref{eq:37})] is independent of the spin indices. Hence, we obtain the spin-symmetric and antisymmetric response by summing those contributions as follows:
\begin{equation}
\frac{{\rm Im} \, G_{T/L}^s({\bf q}, \omega)}{C} = P_{ij}^{L/T} \, {\rm Im} \left(i C_{ij}^{(1)} + i C_{ij}^{(2)} + i C_{ij}^{(3)}\right) \label{eq:38}
\end{equation}
and
\begin{equation}
\frac{{\rm Im} \, G_{T/L}^a({\bf q}, \omega)}{C} = P_{ij}^{L/T} \, {\rm Im} \left(i C_{ij}^{(1)} - i C_{ij}^{(2)}\right), \label{eq:39}
\end{equation}
where $P_{ij}^L = q_i q_j/q^2$ and $P_{ij}^T = (\delta_{ij} - q_i q_j/q^2)/(d-1)$ are the projectors onto the longitudinal and the transverse parts, respectively. Remember that we have to shift $\omega \rightarrow \omega + i \varepsilon$ when evaluating the Wilson coefficients [cf. the definitions in Eqs.~(\ref{eq:6}) and (\ref{eq:7})]. As discussed in the Appendix, of the above terms, only ${\rm Im} \, i C_{ij}^{(3)}$ depends on the scattering length away from the one-particle peak. Barring the dependence of the contact on $a^{-1}$, this means that, except for the spin-symmetric longitudinal response function, all response functions are independent of the scattering length in the asymptotic regime.
 \begin{figure*}[t]
\centering
\subfigure[]{\scalebox{0.53}{\includegraphics[bb = 0 0 453 307]{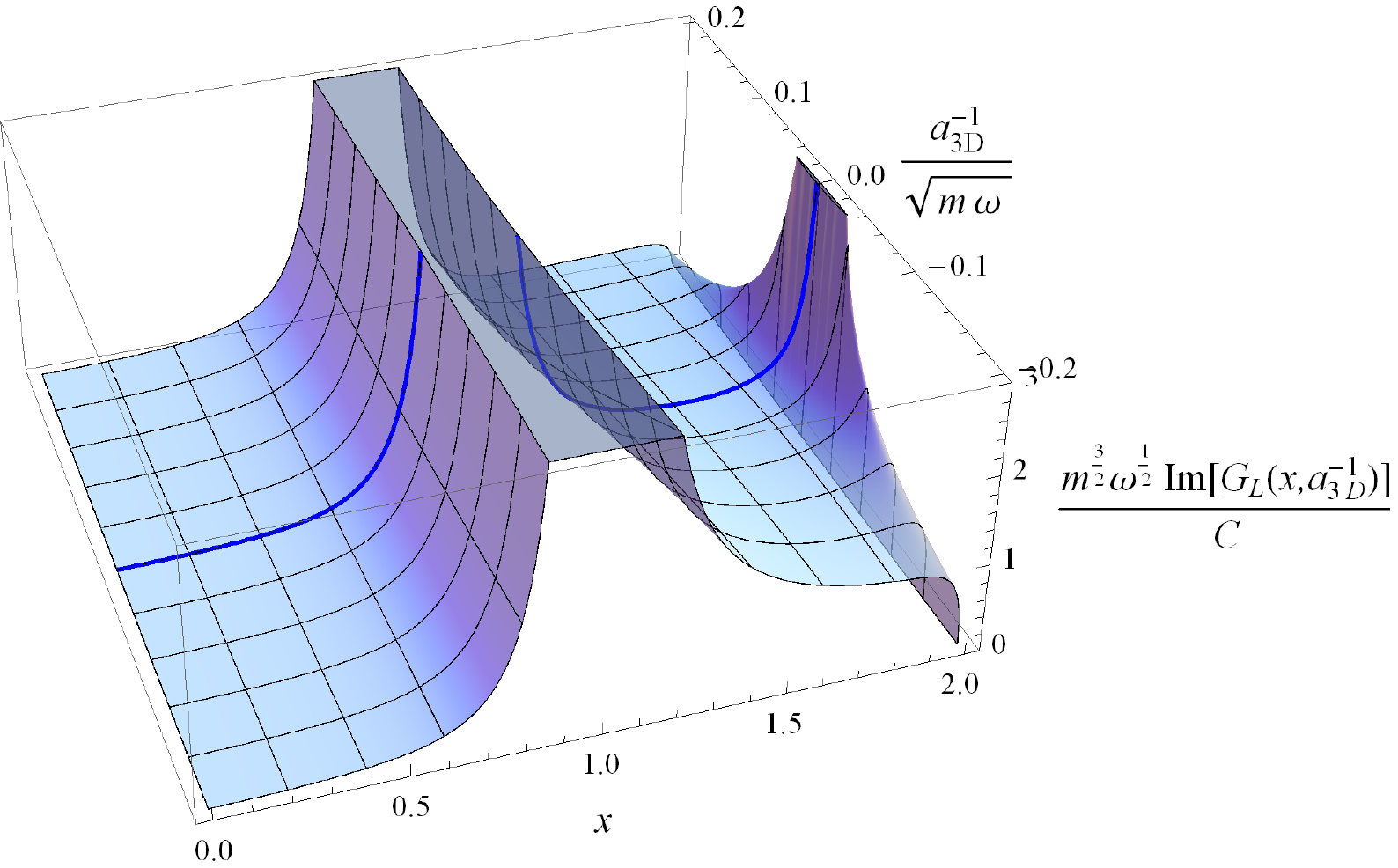}\label{fig:9a}}} \qquad
\subfigure[]{\scalebox{0.53}{\includegraphics[bb = 0 0 453 307]{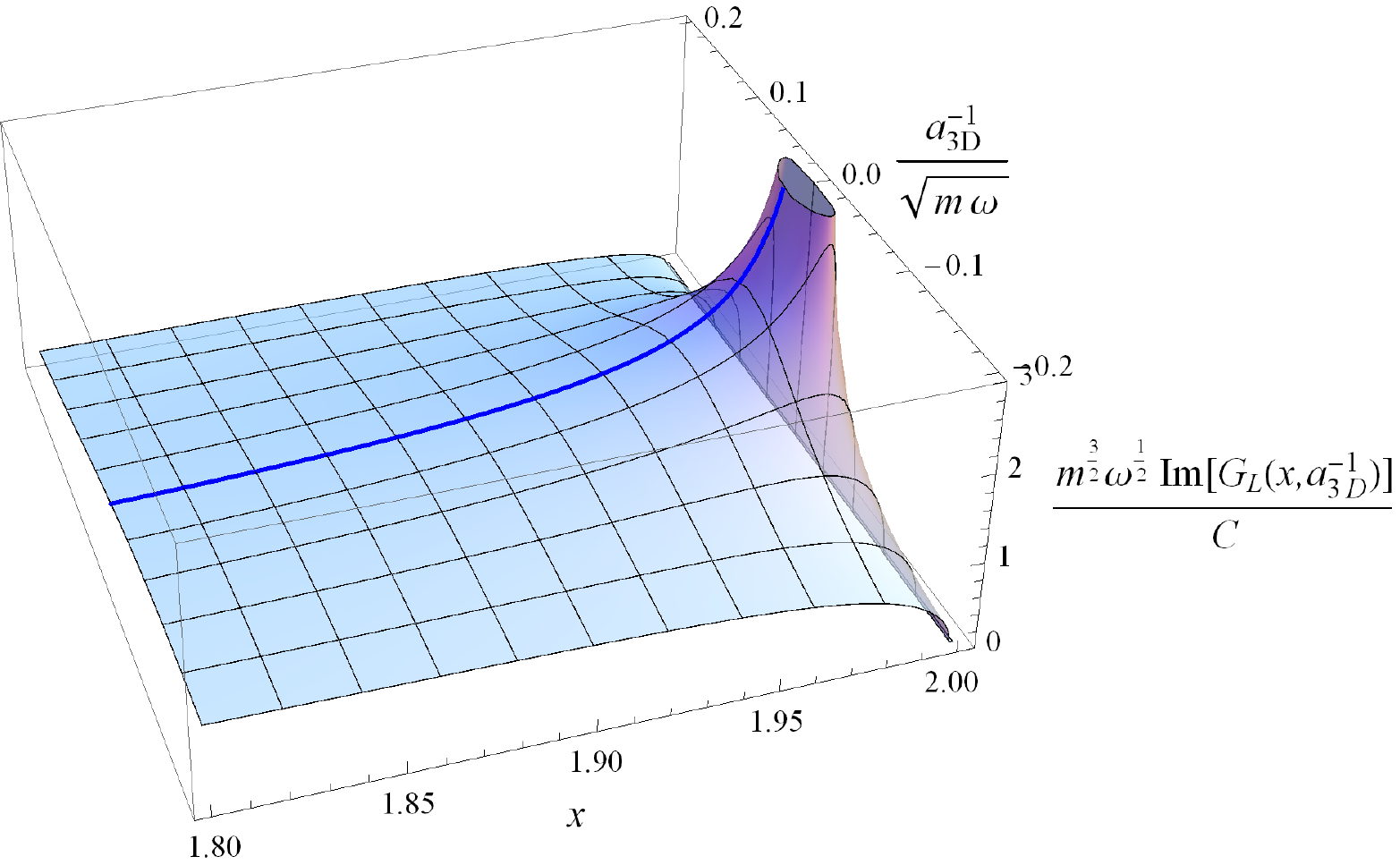}\label{fig:9b}}}
\caption{3D: (a) Longitudinal response function as a function of $x$ and inverse scattering length $a_{\rm 3D}^{-1}/\sqrt{m\omega}$. (b) Detail of (a) at unitarity. The peak at unitarity is due to the strong interaction. The thick blue line denotes the response at unitarity and corresponds to the thick blue line in Fig.~\ref{fig:6a}.}
\label{fig:9}
\end{figure*}
 \begin{figure*}[t]
\centering
\subfigure[]{\includegraphics[scale=0.9]{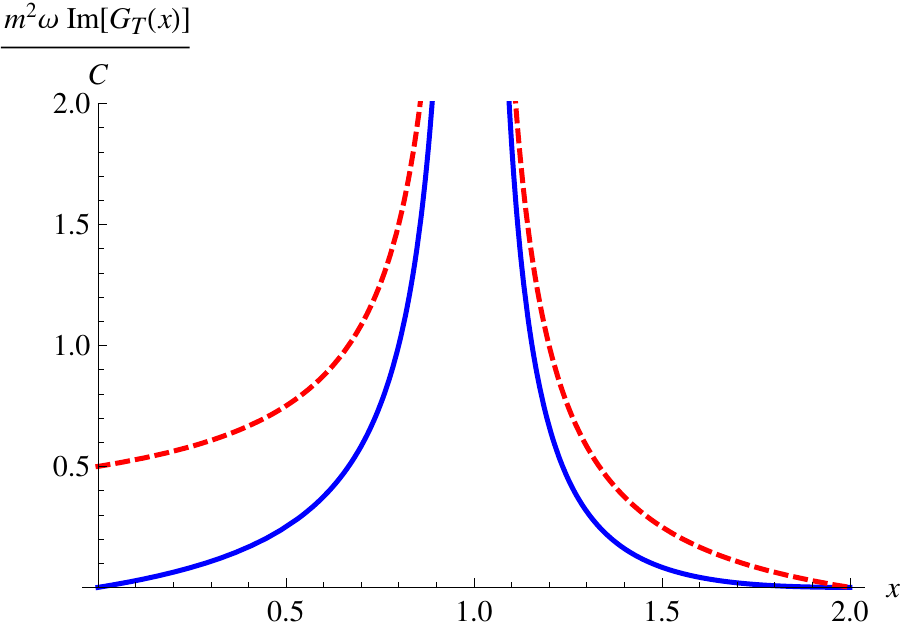}\label{fig:7a}} \qquad
\subfigure[]{\includegraphics[scale=0.9]{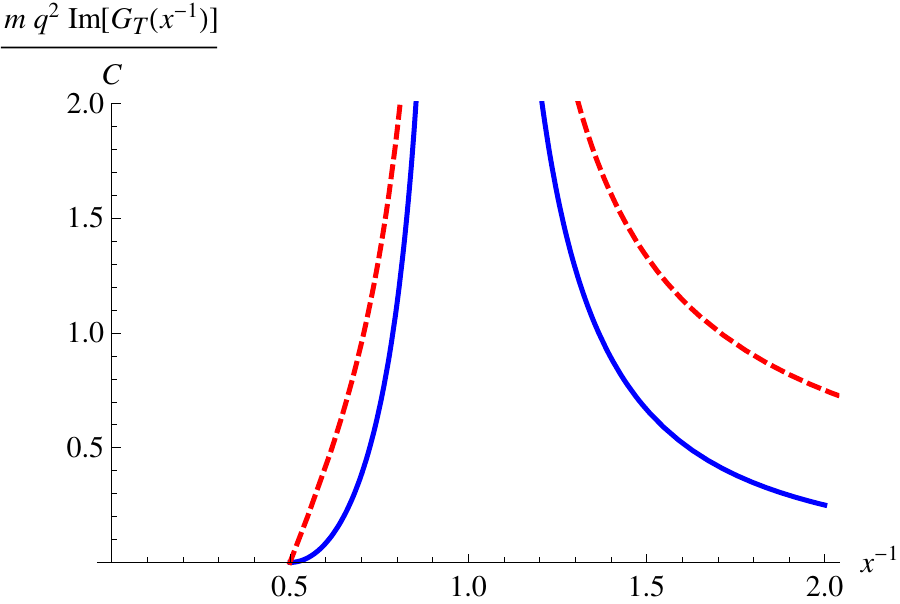}\label{fig:7b}}
\caption{2D: Transverse response function at (a) fixed energy and (b) fixed momentum as a function of the variable $x = q^2/2m\omega$ and $x^{-1}$, respectively. The continuous blue line denotes the spin-symmetric and the red dashed line the spin-antisymmetric response.}
\label{fig:7}
\end{figure*}

The imaginary part of the response functions presented in the next two sections vanishes for $x > 2$. As already pointed out in \cite{12}, this represents the two-particle threshold. This lower bound $\omega > q^2/4m$ follows from energy and momentum conservation if we consider the absorption of a probe with (large) momentum ${\bf q}$ and energy $\omega$ by two particles in the medium.

\subsubsection{3D}

The contributions of Eqs.~(\ref{eq:35})--(\ref{eq:37}) to Eqs.~(\ref{eq:38}) and~(\ref{eq:39}) can be inferred from the results in the Appendix. We start by listing the response functions for the three-dimensional case. The spin-symmetric and antisymmetric current responses are
\begin{align}
&\frac{{\rm Im} \, G_{T}^{s}({\bf q}, \omega)}{C} = - \frac{3}{4 \pi m q^2} \sqrt{m \omega - q^2/4} \nonumber \\
&\quad + \frac{2 (m \omega)^2 + (m \omega - q^2/2)^2}{4 \pi m^2 \omega q^3} \ln \frac{m \omega + q \sqrt{m \omega - q^2/4}}{|m \omega - q^2/2|} \label{eq:40}
\end{align}
and
\begin{align}
&\frac{{\rm Im} \, G_{T}^{a}({\bf q}, \omega)}{C} = - \frac{1}{4 \pi m q^2} \sqrt{m \omega - q^2/4}  \nonumber \\
&\quad + \frac{2 (m \omega)^2 - (m \omega - q^2/2)^2}{4 \pi m^2 \omega q^3} \ln \frac{m \omega + q \sqrt{m \omega - q^2/4}}{|m \omega - q^2/2|} .
\end{align}
Figure~\ref{fig:5} shows plots of the transverse response function as a function of the scaling variable $x= q^2/2m\omega$ [Fig.~\ref{fig:5a}] and $x^{-1}$ [Fig.~\ref{fig:5b}]. They can be interpreted as plots at fixed energy and momentum, respectively. As mentioned earlier, the transverse components do not depend on the scattering length.

\begin{figure*}[t]
\centering
\subfigure[]{\includegraphics[scale=0.9]{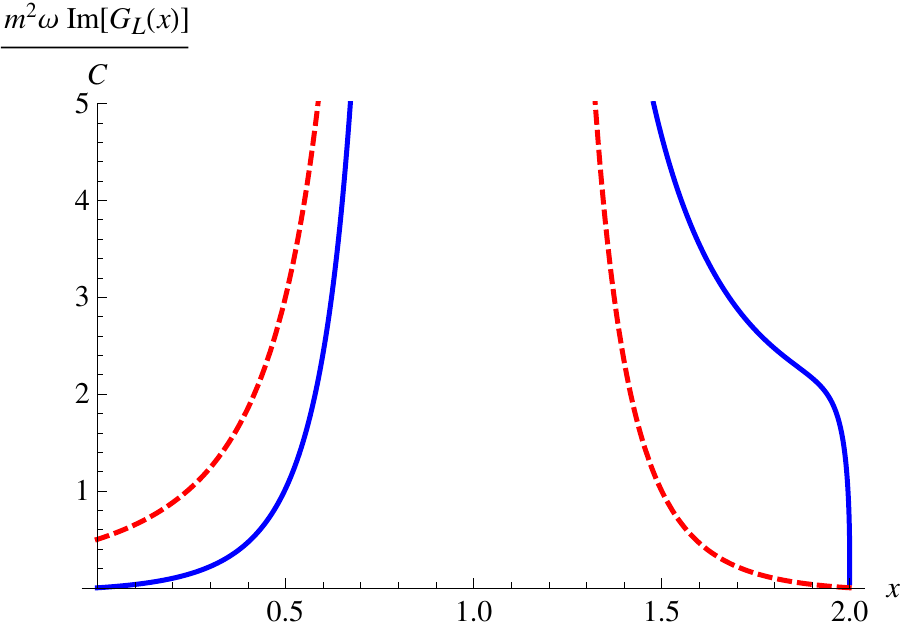}\label{fig:8a}} \qquad
\subfigure[]{\includegraphics[scale=0.9]{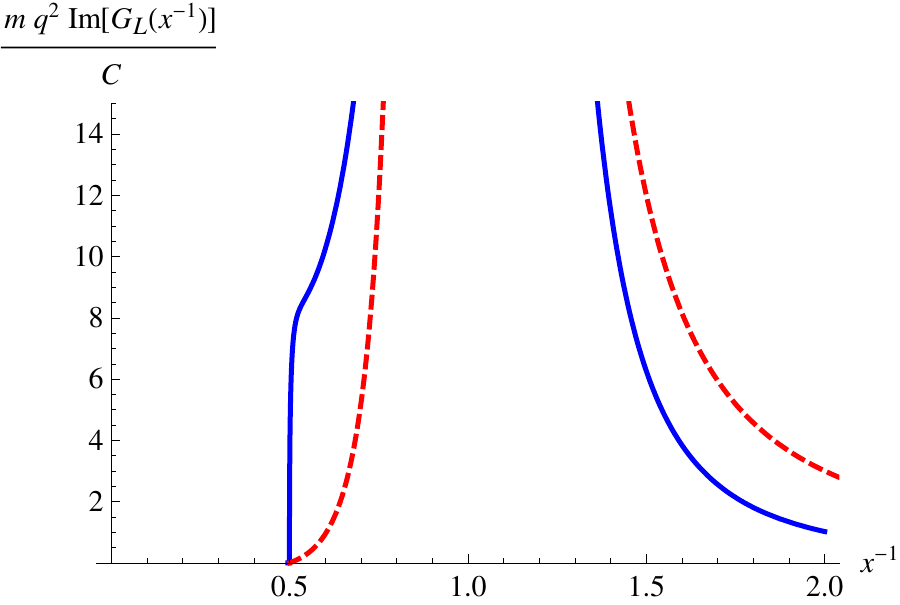}\label{fig:8b}}
\caption{2D: Longitudinal response function at (a) fixed energy and (b) fixed momentum as a function of the variable $x = q^2/2m\omega$ and $x^{-1}$, respectively. The continuous blue line denotes the spin-symmetric and the red dashed line the spin-antisymmetric response. The spin-symmetric response is evaluated at $a_{\rm 2D} \sqrt{m \omega} = 5$ and $a_{\rm 2D} q = 5$.}
\label{fig:8}
\end{figure*}
\begin{figure*}[t]
\centering
\subfigure[]{\scalebox{0.53}{\includegraphics[bb = 0 0 453 307]{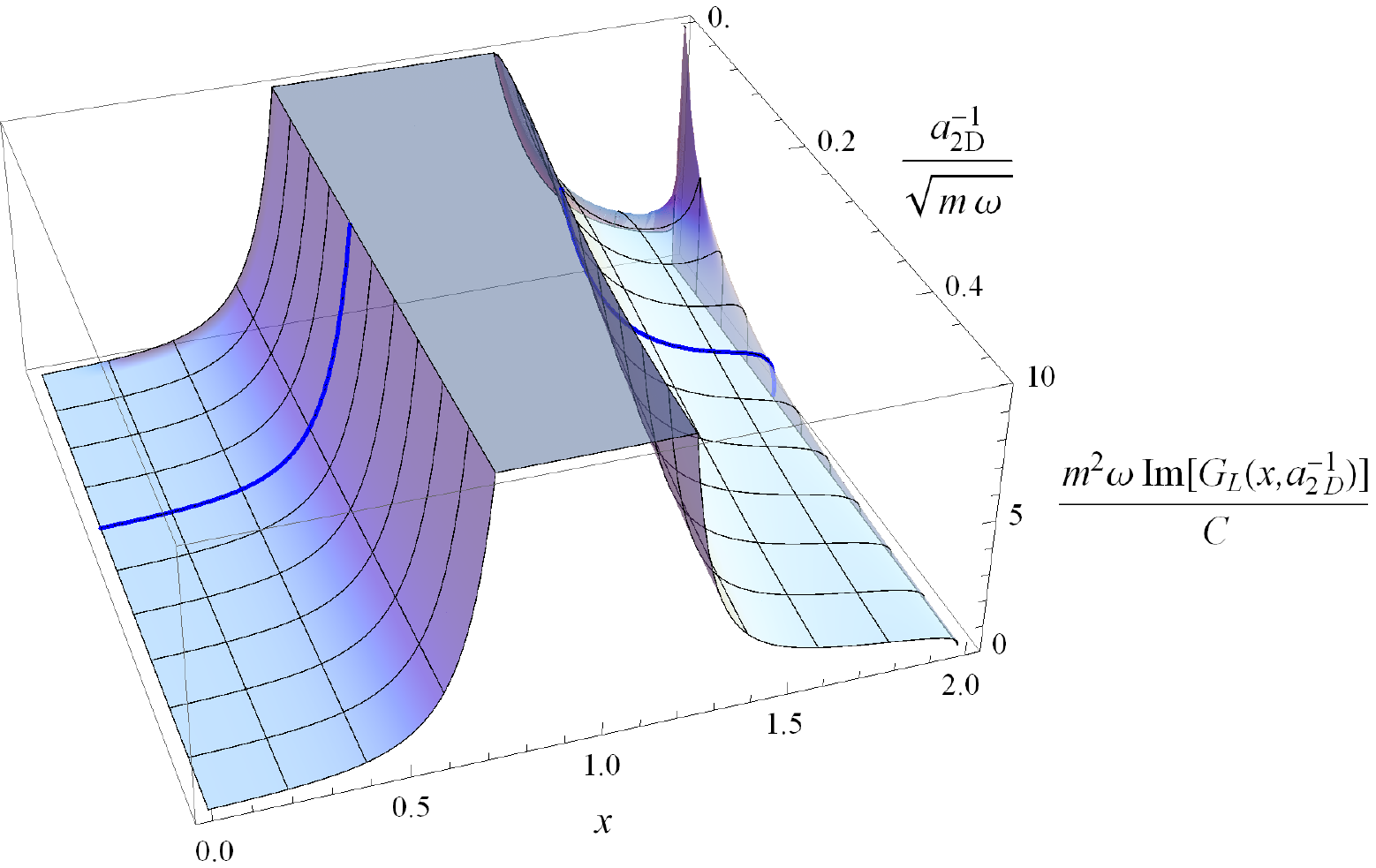}\label{fig:10a}}} \qquad
\subfigure[]{\scalebox{0.53}{\includegraphics{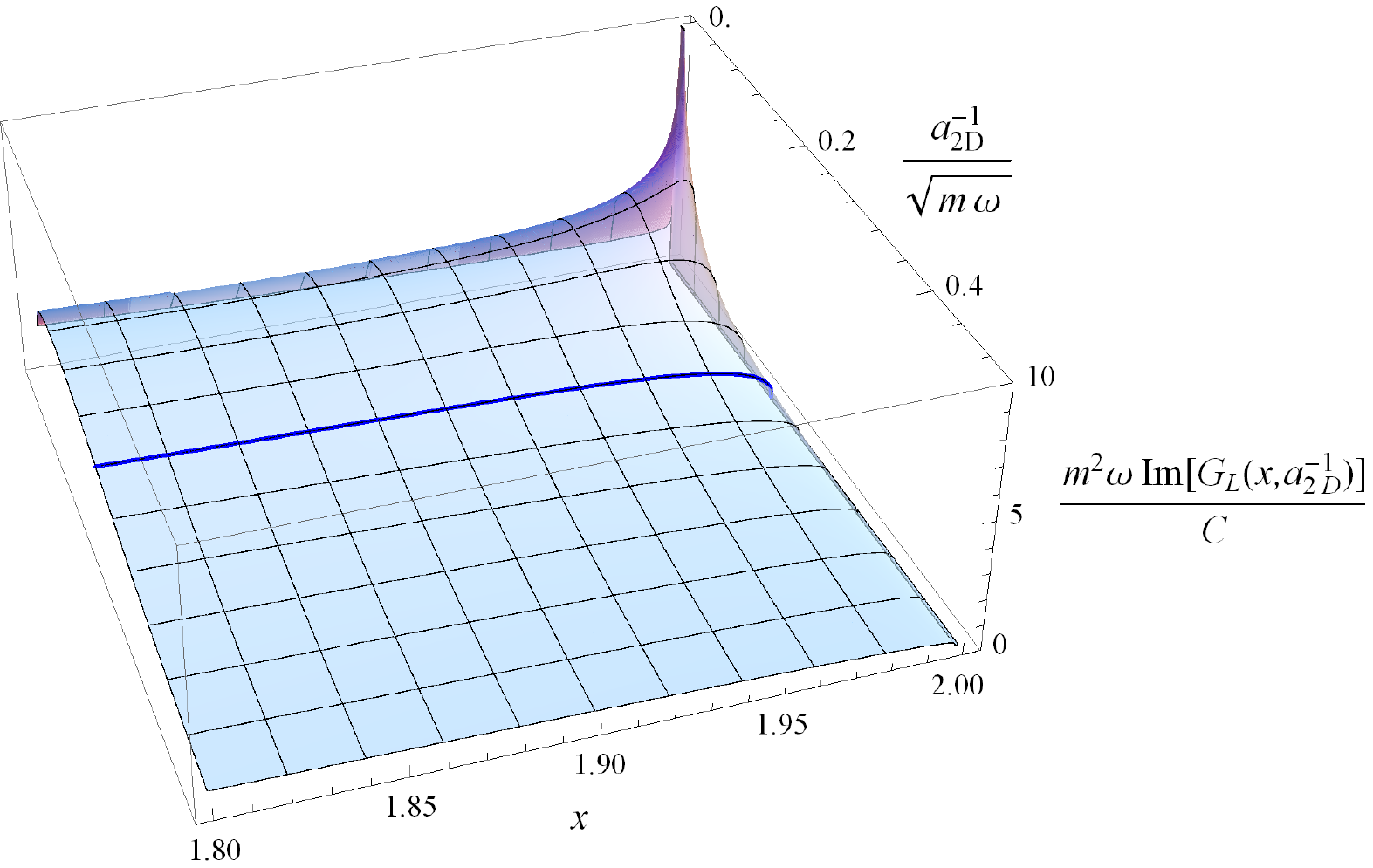}\label{fig:10b}}}
\caption{2D: Longitudinal response function as a function of $x$ and the 2D-scattering length $a_{\rm 2D}^{-1}/\sqrt{m \omega}$. There is a peak at $(a_{\rm 2D}^{-1} = 0, x = 2)$ due to the enhanced interaction in this limit. The thick blue line denotes the response at $a_{\rm 2D}^{-1}/\sqrt{m\omega} = 0.2$ and corresponds to the thick blue line in Fig.~\ref{fig:8a}.}
\label{fig:10}
\end{figure*}
Similarly, we derive the longitudinal response functions
\begin{widetext}
\begin{align}
&\frac{{\rm Im} \, G^{a}_{L}({\bf q}, \omega)}{C} = \frac{m \omega^2}{2 \pi q^2} \frac{\sqrt{m \omega - q^2/4}}{(m \omega - q^2/2)^2} - \frac{\omega}{2 \pi q^3} \ln \frac{m \omega + q \sqrt{m \omega - q^2/4}}{|m \omega - q^2/2|}
\end{align}
and
\begin{align}
\frac{{\rm Im} \, G^{s}_{L}({\bf q}, \omega, a_{\rm 3D}^{-1})}{C} &= \frac{m \omega^2}{2 \pi q^2} \frac{\sqrt{m \omega - q^2/4}}{(m \omega - q^2/2)^2} \left[1 + \frac{2 a_{\rm 3D}^{-1}}{q} \frac{m \omega - q^2/2}{a_{\rm 3D}^{-2} + m \omega - q^2/4} \left(\frac{a_{\rm 3D}^{-1} q}{m \omega - q^2/2} + 2 \pi \, \Theta\left(q^2/2 - m \omega\right)\right) \right] \nonumber \\
&\quad + \frac{\omega}{2 \pi q^3} \ln \frac{m \omega + q \sqrt{m \omega - q^2/4}}{|m \omega - q^2/2|} \left[1 - \frac{4 m \omega a_{\rm 3D}^{-1}/q}{a_{\rm 3D}^{-2} + m \omega - q^2/4} \left(\frac{a_{\rm 3D}^{-1} q}{m \omega - q^2/2} + \pi \, \Theta\left(q^2/2 - m \omega\right)\right)\right] \nonumber \\
&\quad - \frac{m \omega^2}{\pi q^4} \frac{\sqrt{m \omega - q^2/4}}{a_{\rm 3D}^{-2} +m \omega - q^2/4} \, \biggl( \ln^2 \frac{m \omega + q \sqrt{m \omega - q^2/4}}{|m \omega - q^2/2|} - \pi^2 \, \Theta\left(q^2/2 - m \omega\right) \biggr) . \label{eq:43}
\end{align}
At unitarity, this result simplifies considerably:
\begin{align}
\frac{{\rm Im} \, G^{s}_{L}({\bf q}, \omega)}{C} &= \frac{m \omega^2}{2 \pi q^2} \frac{\sqrt{m \omega - q^2/4}}{(m \omega - q^2/2)^2} + \frac{\omega}{2 \pi q^3} \ln \frac{m \omega + q \sqrt{m \omega - q^2/4}}{|m \omega - q^2/2|}  \nonumber \\
&\quad - \frac{m \omega^2}{\pi q^4} \frac{1}{\sqrt{m \omega - q^2/4}} \biggl( \ln^2 \frac{m \omega + q \sqrt{m \omega - q^2/4}}{|m \omega - q^2/2|} - \pi^2 \, \Theta\left(q^2/2 - m \omega\right) \biggr). \label{eq:44}
\end{align}
\end{widetext}
Figure~\ref{fig:6} shows plots of the longitudinal response function as a function of $x$ [Fig.~\ref{fig:6a}] and $x^{-1}$ [Fig.~\ref{fig:6b}], where ${\rm Im} \, G^{s}_{L}({\bf q}, \omega)$ is evaluated at unitarity. In Fig.~\ref{fig:9}, we show plots of the longitudinal response function as a function of both $x$ and the scattering length $a_{\rm 3D}^{-1}/\sqrt{m\omega}$. As it has already been pointed out in~\cite{12}, the peak at $x = 2$ in Fig.~\ref{fig:6a} is due to the strong interactions between the spin species at unitarity. This is also nicely illustrated in Fig.~\ref{fig:9a}, where we find a peak at $a_{\rm 3D}^{-1} = 0$ and a finite response otherwise. A detail of the peak is shown in Fig.~\ref{fig:9b}.

\subsubsection{2D}

For the transverse response function, we find the result
\begin{align}
&\frac{{\rm Im} \, G^{s}_{T}({\bf q}, \omega)}{C} = \frac{q^2}{8 m^2 \omega} \frac{1}{|m \omega - q^2/2|} - \frac{2}{m q^2} \, \Theta\left(\frac{q^2}{2} - m \omega\right) \label{eq:45}
\end{align}
for the spin-symmetric and
\begin{equation}
\frac{{\rm Im} \, G^{a}_{T}({\bf q}, \omega)}{C} = \frac{m \omega - q^2/4}{2 m^2 \omega |m \omega - q^2/2|}
\end{equation}
for the spin-antisymmetric current. Both functions are plotted in Fig.~\ref{fig:7}, as a function of $x$ [Fig.~\ref{fig:7a}] and $x^{-1}$ [Fig.~\ref{fig:7b}], as before.

As for the three-dimensional case, the spin-symmetric longitudinal response function depends on the two-dimensional scattering length $a_{\rm 2D}$:
\begin{align}
&\frac{{\rm Im} \, G^{s}_{L}({\bf q}, \omega, a_{\rm 2D}^{-1})}{C} \nonumber \\
&\quad = \frac{\omega q^2}{8 |m \omega - q^2/2|^3} - \frac{2 m \omega^2}{q^2 (m \omega - q^2/2)^2} \, \Theta\left(\frac{q^2}{2} - m \omega\right)\nonumber \\
&\quad + \frac{1}{\pi^2 + \ln^2 a_{\rm 2D}^2 (m \omega - q^2/4)} \, \frac{4 m \omega^2}{q^2 (m \omega - q^2/2)^2} \nonumber \\
&\quad \times \biggl[\frac{q^2}{4 m \omega} - \ln \frac{m \omega - q^2/4}{|m \omega - q^2/2|} + \Theta\left(\frac{q^2}{2} - m \omega\right) \nonumber \\
&\quad \times \ln a_{\rm 2D}^2 (m \omega - q^2/4)\biggr]^2. \label{eq:47}
\end{align}
The spin-antisymmetric response takes a much simpler form
\begin{align}
&\frac{{\rm Im} \, G^{a}_{L}({\bf q}, \omega)}{C} = \frac{\omega}{2} \frac{m \omega - q^2/4}{|m \omega - q^2/2|^3} .
\end{align}
Figure~\ref{fig:8} shows plots of the longitudinal response function as a function of $x$ [Fig.~\ref{fig:8a}] and $x^{-1}$ [Fig.~\ref{fig:8b}]. Figure~\ref{fig:10} shows a plot of the longitudinal response function as a function of $x$ and the scattering length $a_{\rm 2D}^{-1}/\sqrt{m \omega}$.

It is interesting to note that, for infinite scattering length $a_{\rm 2D}^{-1} = 0$, there is also a peak at $x = 2$, despite the fact that the two-dimensional Fermi gas is not strongly interacting. This reflects the fact that the scattering between the particles is enhanced in this limit, as it can be seen from the two-particle scattering amplitude in Eq.~(\ref{eq:15}). The large value of the scattering length serves to balance the small energy above the two-particle threshold $\omega - q^2/4m$.

When comparing Figs.~\ref{fig:4} and~\ref{fig:6} to~\ref{fig:7} and~\ref{fig:8}, we see that the width of the single-particle peak is much larger in 2D than it is in 3D. In 3D, the decay of the Wilson coefficients away from the pole at $x = 1$ has a logarithmic and a power-law behavior $\sim |1 - x|^{-2}$ for the transverse and the longitudinal responses, respectively, as opposed to higher power-law decays $\sim |1 - x|^{-1}$ and $\sim |1 - x|^{-3}$ in 2D. This is to be expected since we integrate over one space component less in 2D.

\subsection{Dynamic structure factor}\label{sec:3b}

The Ward identity that follows from the U(1) symmetry of the Lagrangian in Eq.~(\ref{eq:1}) gives the first hydrodynamic equation describing the conservation of mass,
\begin{equation}
 \partial_t \langle n \rangle + \nabla \cdot \langle {\bf j} \rangle = 0 , \label{eq:49}
\end{equation}
where ${\bf j} = {\bf j}^\uparrow + {\bf j}^\downarrow$. If we substitute the Fourier transform of Eq.~(\ref{eq:49}) in the spectral decomposition for the imaginary part of the current response function
\begin{align}
&{\rm Im} \, G^{s}_{ij}({\bf q}, \omega) = \pi \left(1 - e^{- \beta \omega}\right) \nonumber \\
&\quad \times \frac{1}{{\cal Z}} \sum_{a,b} e^{- \beta E_a} \langle a | j_i({\bf q}) | b \rangle \langle b | j_j(- {\bf q}) | a \rangle \, \delta(\omega - E_{ba}),
\end{align}
where $E_{ba} = E_b - E_a$ and $E_a$ are the eigenvalues of the Hamiltonian of the system, we can relate the longitudinal part of the spin-symmetric response to the density response $G^\rho$ (where $\rho = m n$ is the mass density) and deduce an expression for the dynamic structure factor
\begin{equation}
S({\bf q}, \omega) = \frac{1}{\pi} {\rm Im} \, G^{\rho}({\bf q}, \omega) = \frac{m^2 q^2}{\pi \omega^2} {\rm Im} \, G^{s}_{L}({\bf q}, \omega) .
\end{equation}
Comparing this with Eqs.~(\ref{eq:43}) and~(\ref{eq:47}), we obtain the asymptotic behavior of the dynamic structure factor. In the three-dimensional case, the result coincides with the expression at unitarity~\cite{12} [cf. Eq.~(\ref{eq:44})]. The general behavior of the structure factor is the same as for the longitudinal response function, Figs.~\ref{fig:9} and~\ref{fig:10}, and we refrain from plotting it again.

When expanding the expression in Eqs.~(\ref{eq:43}) and~(\ref{eq:47}) to leading order in $q$, we get the large frequency tail of the dynamic structure factor for $m q^2 \ll \omega$:
\begin{align}
 &\lim_{\omega \rightarrow \infty} \lim_{{\bf q} \rightarrow 0} S^{{\rm (3D)}}({\bf q}, \omega, a_{\rm 3D}^{-1}) \nonumber \\*
&\quad = \frac{q^4}{\pi^2 m^{1/2} \omega^{7/2}} \left[\frac{4}{45} + \frac{1}{36} \frac{\left(a_{\rm 3D}^{-1}/\sqrt{m \omega}\right)^2}{\left(a_{\rm 3D}^{-1}/\sqrt{m \omega}\right)^2 + 1}\right] C.
\end{align}
At unitarity, this agrees with the result by \cite{14} up to a factor of $2/3$. The same frequency tail was derived by the authors of \cite{12} and \cite{21}, with whom our result agrees. The authors of~\cite{21} also calculated the correction to the tail at leading order in $a_{\rm 3D}^{-1}$. Again, our result is in agreement with theirs. Our calculation generalizes the cited results to finite scattering length. For the two-dimensional Fermi gas, we find no dependence on the scattering length in the high-frequency tail
\begin{equation}
\lim_{\omega \rightarrow \infty} \lim_{{\bf q} \rightarrow 0} S^{{\rm (2D)}}({\bf q}, \omega) = \frac{q^4}{8 \pi m \omega^4} C.
\end{equation}

\subsection{Spectral viscosities}\label{sec:3c}

The authors of~\cite{14} derived Kubo formulas for frequency-dependent generalizations of the bulk and the shear viscosity in three dimensions. The starting point of their derivation is the Euler equation,
\begin{equation}
m \partial_t j_k({\bf x}, t) = - \partial_j \Pi_{jk}({\bf x}, t) ,
\end{equation}
where $\Pi_{ij}$ is the stress tensor for a viscous fluid in $d$ space dimensions:
\begin{equation}
 \Pi_{ij} = p \delta_{ij} + \rho v_i v_j - \sigma'_{ij} .
\end{equation}
$p$ denotes the pressure, ${\bf v}$ the velocity, $\rho$ the density, and
\begin{equation}
 \sigma'_{ij} = \eta \left[\partial_i v_j + \partial_j v_i - \frac{2}{d} \delta_{ij} (\nabla \cdot {\bf v})\right] + \zeta \delta_{ij} (\nabla \cdot {\bf v})
\end{equation}
is the viscous stress tensor \cite{33}. Retracing the derivation in~\cite{14}, it is not difficult to generalize their result to two space dimensions. We obtain the relations
\begin{equation}
{\rm Re} \, \eta(\omega) = \lim_{q \rightarrow 0} \frac{m^2 \omega}{q^2} \, {\rm Im} \, G^{s}_{T}({\bf q}, \omega)
\end{equation}
and
\begin{equation}
\left(2 - \frac{2}{d}\right) {\rm Re} \, \eta(\omega) + {\rm Re} \, \zeta(\omega) = \lim_{q \rightarrow 0} \frac{m^2 \omega}{q^2} \, {\rm Im} \, G^{s}_{L}({\bf q}, \omega).
\end{equation}
The leading order of $\eta(\omega)$ and $\zeta(\omega)$ at large frequency is proportional to the contact. Comparing this with Eqs.~(\ref{eq:40}), (\ref{eq:43}), (\ref{eq:45}), and~(\ref{eq:47}), we obtain the large-frequency tails of the bulk and the shear viscosity:
\begin{equation}
  {\rm Re} \, \eta(\omega) = \frac{C}{15 \pi \sqrt{m \omega}}\label{eq:59}
\end{equation}
and
\begin{equation}
 {\rm Re} \, \zeta(\omega) = \frac{C}{36 \pi \sqrt{m \omega}} \, \frac{\left(a_{\rm 3D}^{-1}/\sqrt{m \omega}\right)^2}{\left(a_{\rm 3D}^{-1}/\sqrt{m \omega}\right)^2 + 1}\label{eq:60}
\end{equation}
in three dimensions, and
\begin{equation}
 {\rm Re} \, \eta(\omega) = \frac{C}{8 m \omega}\label{eq:61}
\end{equation}
and
\begin{equation}
 {\rm Re} \, \zeta(\omega) = 0\label{eq:62}
\end{equation}
in two dimensions. The high-frequency tails have been considered before for a three-dimensional gas near unitarity \cite{14,32}. Equation~(\ref{eq:59}) agrees with \cite{32} and agrees with \cite{14} up to the same factor of $2/3$ mentioned in the previous section. We obtain the high-frequency tail of the 3D bulk viscosity~(\ref{eq:60}) for an arbitrary value of the scattering length. It vanishes at unitarity, in agreement with~\cite{14}. 

\subsection{Viscosity sum rules}

By integrating over the frequency of the spectral functions, the asymptotic behavior of the correlation functions can be used to derive sum rules. On the one hand, such sum rules provide constraints that serve as a consistency check for approximate methods \cite{32}. On the other hand, they are often more closely related to experimental results since it can be difficult to probe the strictly asymptotic region of the response functions. The simplest way to obtain sum rules is to calculate the lowest momenta of the spectral functions. Additional sum rules can be deduced by introducing some form of UV cutoff that shifts the integration region to lower frequencies \cite{11, 21}.

For completeness, we calculate the lowest moment of the bulk and the shear viscosity and rederive the results of Ref.~\cite{14} using the OPE. To obtain a finite expression for the shear viscosity sum rule, we subtract the high-frequency tail in Eq.~(\ref{eq:59}) from the integrand. This yields
\begin{equation}
 \frac{2}{\pi} \int_0^\infty {\rm d}\omega \, \left[\eta(\omega) - \frac{C}{15 \pi \sqrt{m \omega}}\right] = \frac{2}{3} \, H - \frac{a_{\rm 3D}^{-1}}{6 \pi m} \, C \label{eq:63}
\end{equation}
and
\begin{align}
\frac{2}{\pi} \int_0^\infty {\rm d}\omega \, \zeta(\omega) &= \frac{10}{9} \, H \nonumber + \frac{2 a_{\rm 3D}^{-1}}{9 \pi m} \, C \\
&\quad + \lim_{\omega \rightarrow 0} \lim_{q \rightarrow 0} \frac{m^2 \omega^2}{q^2} \, {\rm Im} \, G_{L}^{s}({\bf q}, \omega) \, \label{eq:64}.
\end{align}
The shear viscosity sum rule~(\ref{eq:63}) agrees with the result in \cite{32} and agrees with \cite{14} up to the prefactors.

The OPE alone is not sufficient to constrain the bulk viscosity sum rule~(\ref{eq:64}). As it was pointed out in Ref.~\cite{14}, this is due to an ordering ambiguity when taking the zero-momentum limit ${\bf q} \rightarrow 0$ and performing the $\omega$ integration (we take the limit \emph{after} performing the integration). The term in the second line of Eq.~(\ref{eq:64}) is evaluated in the zero-momentum and zero-frequency limit where the OPE cannot provide an insight. It was evaluated explicitly in~\cite{14} using hydrodynamic expressions for the current response. Combining the result in~\cite{14} with Eq.~(\ref{eq:64}) yields the sum rule
\begin{align}
&\frac{2}{\pi} \int_0^\infty {\rm d}\omega \, \zeta(\omega) = \frac{1}{36 \pi m}\left.\frac{\partial (a_{\rm 3D}^{-2} C)}{\partial a_{\rm 3D}^{-1}}\right|_s,
\end{align}
where the subscript $s$ indicates that we take the derivative of $C$ with respect to $a_{\rm 3D}^{-1}$ at fixed entropy.

\section{SUMMARY}\label{sec:4}

In this paper, we studied the current response function of a Fermi gas and calculated the first terms of its operator-product expansion. We considered the limit $\omega, {\bf q} \rightarrow \infty$ away from the one-particle peak $\omega \neq q^2/2m$ with $q^2/m\omega$ and $a^{-1}/\sqrt{m \omega}$ held fixed. We found that, to leading order, the response is proportional to the contact whose Wilson coefficient we determined. We calculated various response functions for a wide range of parameters: we derived the longitudinal and transverse components of the spin-symmetric and antisymmetric response functions in 3D and 2D for arbitrary values of the scattering length.

Using those results, we were able to obtain the asymptotic form of the dynamic structure factor in 3D and 2D and thus to generalize and to extend many of the previous results for 3D systems at large scattering length \cite{12,21,14}. We rederived the high-frequency tails of the bulk and shear viscosity considered in \cite{14} using the OPE and calculated their 2D counterparts. Furthermore, we could use the OPE to calculate some sum rules that constrain the spectral viscosities in 3D.

As illustrated by this work, the operator-product expansion provides a powerful tool to derive universal relations. It offers a systematic way to analyze the short-time and distance structure of a theory, which can be fully characterized by performing calculations involving few-body states.

\emph{Note added.} --- Recently, Ref.~\cite{36} appeared. The authors calculate the viscosity tails in 3D and obtain a result in agreement with Eq.~(\ref{eq:59}) and~(\ref{eq:60}) of this paper.

\section*{Acknowledgments}

I would like to thank Matthew Wingate for helpful discussions and for reading the manuscript. I am supported by CHESS, DAAD, STFC, St.~John's College, Cambridge, and by the Studienstiftung des deutschen Volkes.

\appendix

\section{MATRIX ELEMENTS FOR THE TWO-PARTICLE STATE}\label{sec:a}

This appendix gives some details of the calculation of the Feynman diagrams in Fig.~\ref{fig:4}. We begin by sketching the most important steps in the evaluation of the matrix elements between the two-particle state. Section~\ref{sec:a1} gives an example of such a calculation and shows how to derive the contribution $C_{ij}^{(3)}$ to the Wilson coefficient of the contact operator [Eq.~(\ref{eq:37})]. The results for the remaining contributions $C_{ij}^{(1)}$ and $C_{ij}^{(2)}$ are listed in Sec.~\ref{sec:a2}.

We evaluate each diagrams using the Feynman rules summarized in Sec.~\ref{sec:2a}. For all diagrams involving an internal loop with loop momentum ${\bf k}$ and energy $k_0$, the $k_0$ integration can be carried out trivially using the theorem of residues, which gives the expressions in Eqs.~(\ref{eq:35}),~(\ref{eq:36}), and~(\ref{eq:37}). In the next step, we perform a partial fraction decomposition of the integrand with respect to ${\bf k}$. Where necessary, we introduce Feynman parameters and shift the integration variable to eliminate the angular dependence in the denominator. Since the resulting expression is rotationally invariant, we only need to retain the terms that are symmetric under ${\bf k} \rightarrow - {\bf k}$. Rotational invariance also dictates the tensorial structure of the integrals. That way, we can express the integral as a sum containing (scalar) integrals of the type
\begin{equation}                  
 \int \frac{{\rm d}^d {\bf k}}{(2 \pi)^d} \, \frac{k^a}{(k^2 - \alpha^2 - i \varepsilon)^b},
\end{equation}
where $\alpha^2$ is some constant (that typically depends on $\omega$, ${\bf q}$, and on a Feynman parameter). The value of those integrals can be determined by performing the integral for arbitrary $a$ and $b = 1$ and differentiating the results with respect to $\alpha^2$. Of course, one can also analytically continue $\alpha \rightarrow i \alpha$ and evaluate the resulting integral.
It is convenient to introduce IR and UV cutoffs $\eta$ and $\Lambda$, respectively, when performing the calculations. It remains to calculate the integrals over the Feynman parameter. They take the form
\begin{equation}
 \int_0^1 {\rm d}x \, \frac{(1 - x)^b}{[x (m \omega - q^2/2) + x^2 q^2/4]^a}. \label{eq:a2}
\end{equation}
If we define the variable $z = q^2/4m\omega$ and perform the transformations $z \rightarrow z' = z/(1-z)$ and then $x \rightarrow 1 - x$, we can recast Eq.~(\ref{eq:a2}) in the form
\begin{align}
&\frac{1}{(m \omega - q^2/4)^a} \int_0^1 {\rm d}x \, \frac{x^b (1 - x)^{-a}}{[1 - x z']^a} \nonumber \\
&\quad = \frac{1}{(m \omega - q^2/4)^a} \frac{\Gamma(1 + b) \Gamma(1 - a)}{\Gamma(2 - a + b)} \nonumber \\
&\quad \quad \times {}_2F_1\left(a, b+1; 2 - a + b; z'\right), \label{eq:a3}
\end{align}                                                                          
where the left-hand side defines the Gaussian hypergeometric function ${}_2F_1$ \cite{34}. The functions can be further simplified using standard identities such as \cite{35}. We can rewrite the resulting expression in terms of elementary functions using any computer algebra system. In 2D, additional integrals may arise where the integrand depends logarithmically on $x$. They can be solved with the aid of a computer algebra system as well.

While the evaluation of each Feynman diagram proceeds as outlined above, the calculations can be very cumbersome and will not be reproduced here. As an example, we consider only the simplest diagrams and calculate the matrix elements of the three diagrams in Fig.~\ref{fig:4c} in the next section. The results of the other diagrams are given in Sec.~\ref{sec:a2}

\subsection{The contribution $C_{ij}^{(3)}$ to the contact's Wilson coefficient}\label{sec:a1}

To illustrate the procedure outlined in the previous section, we calculate the sum of the diagrams in Fig.~\ref{fig:4c}. If we apply the Feynman rules in Sec.~\ref{sec:2a} to evaluate Fig.~\ref{fig:4c}, we obtain
\begin{align}
 &C_{ij}^{(3)} = - \frac{i {\cal A}(\omega, {\bf q})}{m^4} \biggl[ \frac{{\cal A}^{-1}}{\omega - E_{\bf q}} q_i + \int \frac{{\rm d}k_0}{2 \pi} \int \frac{{\rm d}^d {\bf k}}{(2 \pi)^d} \nonumber \\
&\quad \times \frac{i}{k_0 - E_{\bf k} + i \varepsilon} \frac{i}{k_0 + \omega - E_{\bf k + q} + i \varepsilon} \nonumber \\
&\quad \times \frac{i}{- k_0 - E_{\bf k} + i \varepsilon} (2 k_i + q_i)\biggr] \times [i \rightarrow j] + (\omega \rightarrow - \omega), \label{eq:a4}
\end{align}
where we divided by the matrix element of the contact operator in order to obtain the contribution $C_{ij}^{(3)}$ to the contact's Wilson coefficient. Equation~(\ref{eq:a4}) can be further evaluated using the theorem of residues
\begin{align}
 C_{ij}^{(3)} &= - \frac{i {\cal A}(\omega, {\bf q})}{m^4} \biggl[ \frac{q_i}{\omega - E_{\bf q}} {\cal A}^{-1} + \int \frac{{\rm d}^d {\bf k}}{(2 \pi)^d} \nonumber \\
&\quad\times \frac{1}{2 E_{\bf k}} \frac{1}{\omega - E_{\bf k + q} - E_{\bf k} + i \varepsilon} (2 k_i + q_i)\biggr] \nonumber \\
&\quad \times [i \rightarrow j] + (\omega \rightarrow - \omega),
\end{align}
which coincides with the expression in Eq.~(\ref{eq:37}).

In the next step, we introduce Feynman parameters and shift the integration variable ${\bf k} \rightarrow {\bf k} - x {\bf q}/2$ to obtain a rotationally invariant integrand. The term $\sim k_i$ is odd under ${\bf k} \rightarrow - {\bf k}$ and can be dropped. This yields
\begin{align}
 &C_{ij}^{(3)} = - i {\cal A}(\omega, {\bf q}) \biggl[ \frac{1}{m \omega - {\bf q}^2/2} \frac{1}{m {\cal A}} - \int_0^1 {\rm d}x (1 - x) \nonumber \\
& \times \int \frac{{\rm d}^d {\bf k}}{(2 \pi)^d} \frac{1}{\left\{k^2 - \left(x (m \omega - q^2/2) + x^2 q^2/4\right) - i \varepsilon\right\}^2}\biggr]^2 q_i q_j \nonumber \\
&+ (\omega \rightarrow - \omega). \label{eq:a6}
\end{align}
In order to perform the ${\bf k}$ integration, we note the following identity:
\begin{equation}
\int \frac{{\rm d}^d {\bf k}}{(2 \pi)^d} \frac{1}{k^2 - \alpha^2 - i \varepsilon} = \left\{\begin{matrix} \frac{1}{2 \pi^2} \left(\Lambda + \frac{i \pi}{2} \alpha\right) & d = 3 \\ \frac{1}{2 \pi} \left(\frac{1}{2} \ln \frac{\Lambda^2}{\alpha^2} + \frac{i \pi}{2}\right) & d = 2 \end{matrix} \right. .
\end{equation}
Differentiating this with respect to $\alpha^2$ gives the desired result
\begin{equation}
\int \frac{{\rm d}^d {\bf k}}{(2 \pi)^d} \frac{1}{[k^2 - \alpha^2 - i \varepsilon]^2} = \left\{\begin{matrix} \frac{i}{8 \pi \alpha} & d = 3 \\ - \frac{1}{4 \pi \alpha^2} & d = 2 \end{matrix} \right. , \label{eq:a8}
\end{equation}
where we have to put $\alpha^2 = x (m \omega - q^2/2) + x^2 q^2/4$.
We consider the 3D case first. Inserting Eq.~(\ref{eq:a8}) in~(\ref{eq:a6}) and using~(\ref{eq:a3}) gives
\begin{align}
 &C_{ij}^{(3)} = - i {\cal A}(\omega, {\bf q}) \biggl[ \frac{1}{m \omega - {\bf q}^2/2} \frac{1}{m {\cal A}}  \nonumber \\
&- \frac{i}{6 \pi} \frac{1}{\sqrt{m \omega - q^2/4}} \, {}_2F_1\left(\frac{1}{2}, 2; \frac{5}{2}; z'\right)\biggr]^2 q_i q_j + (\omega \rightarrow - \omega). \label{eq:a9}
\end{align}
The hypergeometric function can be written in terms of elementary functions as follows:
\begin{align}
&{}_2F_1\left(\frac{1}{2}, 2; \frac{5}{2}; z'\right) = - \frac{3}{q^2} \left(m \omega - q^2/4\right) \nonumber \\
&\quad+ \frac{3 m \omega}{q^3} \sqrt{m \omega - q^2/4} \left(\ln \frac{m \omega + q \sqrt{m \omega - q^2/4}}{m \omega - q^2/2}\right) .
\end{align}
Substituting this in Eq.~(\ref{eq:a9}) yields the final result for the Wilson coefficient
\begin{align}
 &C_{ij}^{(3)} = \frac{i {\cal A}(\omega, {\bf q})}{4 \pi^2} \bigg[\frac{i a_{\rm 3D}^{-1}}{2 (m \omega - q^2/2)} + \frac{1}{q^2} \sqrt{m \omega - q^2/4} \nonumber \\
&\quad - \frac{m \omega}{q^3} \, \ln \frac{m \omega + q \sqrt{m \omega - q^2/4}}{m \omega - q^2/2} \bigg]^2 q_i q_j + (\omega \rightarrow - \omega) .
\end{align}
Similarly, we obtain for the 2D case:
\begin{align}
 &C_{ij}^{(3)} = - \frac{i}{4 \pi^2} \, \frac{{\cal A}(\omega, {\bf q})}{(m \omega - q^2/2)^2} \bigg[\ln a_{\rm 2D} \sqrt{- (m \omega - q^2/4)} \nonumber \\
&\quad + \frac{1}{2} - \frac{2 m \omega}{q^2} \ln \frac{m \omega - q^2/4}{m \omega - q^2/2}\bigg]^2 q_i q_j + (\omega \rightarrow - \omega) .
\end{align}

\subsection{List of contributions to the contact's Wilson coefficient}\label{sec:a2}

For completeness, we state the contributions of the other diagrams in Fig.~\ref{fig:4} to the Wilson coefficients without giving an explicit derivation. Figure~\ref{fig:4a} contributes a term
\begin{widetext}
\begin{align}
C_{ij}^{(1)} &=\frac{1}{4 \pi m q^2} \, \biggl(\frac{\sqrt{m \omega - q^2/4}}{(m \omega - q^2/2)^2} \left(q^4 - 4 m \omega q^2 + 6 m^2 \omega^2\right) -  \frac{6 m \omega}{q} \ln \frac{m \omega + q \sqrt{m \omega - q^2/4}}{m \omega - q^2/2} \biggr) \frac{q_i q_j}{q^2} \nonumber \\
&\quad + \frac{1}{2 \pi m q^2} \, \biggl(- \sqrt{m \omega - q^2/4} + \frac{m \omega}{q} \ln \frac{m \omega + q \sqrt{m \omega - q^2/4}}{m \omega - q^2/2}\biggr) \delta_{ij} - c_{ij}^{\cal H}({\bf q}, \omega) \frac{a_{\rm 3D}^{-1}}{4 \pi m} + c_{ij}^{(4)}({\bf q}, \omega) \frac{a_{\rm 3D}^{-1}}{2 \pi m} + (\omega \rightarrow - \omega) 
\end{align}
in 3D. $c_{ij}^{\cal H}$ is defined in Eq.~(\ref{eq:29}) and $c_{ij}^{(4)}$ in Eq.~(\ref{eq:26}). In 2D, we find
\begin{align}
C_{ij}^{(1)} &= \frac{- i}{4 \pi m} \frac{1}{m \omega - q^2/2} \biggl(\frac{42 q^2 m \omega - 7 q^4 - 48 (m \omega)^2}{48 (m \omega - q^2/2)^2} + \frac{4 m \omega}{q^2} \ln \frac{m \omega - q^2/4}{m \omega - q^2/2} - 2 \ln a_{\rm 2D} \sqrt{- (m \omega - q^2/4)}\biggr) \delta_{ij} \nonumber \\
&\quad + \frac{i}{4 \pi m} \frac{q^2}{(m \omega - q^2/2)^3} \biggl(\frac{7 m \omega}{4} - \frac{11 q^2}{24} - \frac{2 m^2 \omega^2}{q^2} + \left(3 m \omega - q^2\right) \ln a_{\rm 2D} \sqrt{- (m \omega - q^2/4)} \nonumber \\
&\quad + \frac{m \omega (8 m^2 \omega^2 - 12 m \omega q^2 + 3 q^4)}{q^4} \ln \frac{m \omega - q^2/4}{m \omega - q^2/2}\biggr) \frac{q_i q_j}{q^2} + (\omega \rightarrow - \omega). 
\end{align} 
Similarly, we obtain for the sum of the diagrams in Fig.~\ref{fig:4b}:
\begin{align}
C_{ij}^{(2)} &= - \frac{1}{4 \pi m} \, \biggl\{\frac{1}{q^2} \sqrt{m \omega - q^2/4} + \frac{m \omega}{q^3} \left(\frac{q^2}{(m \omega)^2} \left[m \omega - \frac{q^2}{4}\right) - 1\right] \ln \frac{m \omega + q \sqrt{m \omega - q^2/4}}{m \omega - q^2/2}\biggr\} \delta_{ij} \nonumber \\
&\quad - \frac{1}{4 \pi m^2 \omega q} \, \biggl[\left(\frac{3 (m \omega)^2}{q^2} - m \omega + \frac{q^2}{4}\right) \, \ln \frac{m \omega + q \sqrt{m \omega - q^2/4}}{m \omega - q^2/2} - \frac{3 m \omega}{q} \sqrt{m \omega - q^2/4}\biggr] \frac{q_i q_j}{q^2} \nonumber \\
&\quad + \frac{i a_{\rm 3D}^{-1}}{8 \pi m^2 \omega} \frac{q^2}{m \omega - q^2/2} \frac{q_i q_j}{q^2} + (\omega \rightarrow - \omega) 
\end{align}
in 3D, and
\begin{align}
C_{ij}^{(2)} &= \frac{i}{4 \pi m^2 \omega} \frac{q^2}{m \omega - q^2/2} \left[\frac{8 m^2 \omega^2 - 4 m \omega q^2 + q^4}{q^4} \ln \frac{m \omega - q^2/4}{m \omega - q^2/2} - \ln a_{\rm 2D} \sqrt{- (m \omega - q^2/4)} - \frac{2}{q^2} \left(m \omega - \frac{q^2}{4}\right) \right] \frac{q_i q_j}{q^2} \nonumber \\         
&\quad+ \frac{i}{2 \pi m^2 \omega} \biggl( \ln \frac{\Lambda}{\sqrt{- (m \omega - q^2/4)}} + \frac{1}{2} - \frac{2}{q^2} \left(m \omega - \frac{q^2}{2}\right) \ln \frac{m \omega - q^2/4}{m \omega - q^2/2} \biggr) \delta_{ij}\ + (\omega \rightarrow - \omega)
\end{align}
in 2D.
\end{widetext}

\end{document}